\documentclass[conference]{IEEEtran}
\usepackage{booktabs} 

\usepackage{mdwmath}
\usepackage{mdwtab}
\usepackage[tight,footnotesize]{subfigure}
\usepackage{amssymb}
\usepackage{algorithm}
\usepackage{graphicx}
\usepackage{graphics}
\usepackage{multirow}
\usepackage{mathtools}
\newcommand{\nop}[1]{}

\begin{document}
\title{SENATUS: An Approach to Joint Traffic Anomaly Detection and Root Cause Analysis }

\author{Atef~Abdelkefi$^1$, Yuming~Jiang$^1$, Sachin Sharma$^2$ \\
{$^1$NTNU - Norwegian University of Science and Technology, Norway. $^2$NEC Laboratories Europe, Germany} 
}

\maketitle

\begin{abstract}
In this paper, we propose a novel approach, called SENATUS, for joint traffic anomaly detection and root-cause analysis. Inspired from the concept of a senate, the key idea of the proposed approach is divided into three stages: election, voting and decision. At the election stage, a small number of \nop{traffic flow sets (termed as senator flows)}senator flows are chosen\nop{, which are used} to represent approximately the total (usually huge) set of traffic flows. In the voting stage, anomaly detection is applied on the senator flows and the detected anomalies are correlated to identify the most possible anomalous time bins. Finally in the decision stage, a machine learning technique is applied to the senator flows of each anomalous time bin to find the root cause of the anomalies. We evaluate SENATUS using traffic traces collected from the Pan European network, GEANT, and compare against another approach which detects anomalies using lossless compression of traffic histograms. We show the effectiveness of SENATUS in diagnosing anomaly types: network scans and DoS/DDoS attacks.
\end{abstract}


\section{Introduction}
New applications, \nop{which emerge}emerging every year or even every day, have made it imperative to investigate effective techniques that can extract communication patterns from Internet traffic for security management\nop{ of the network}. Among others, identifying anomalous events such as denial-of-service (DoS) attacks, distributed DoS (DDoS) attacks and network scans is a crucial task. 

A key challenge in traffic anomaly detection is the curse of dimensionality, which refers to the problems that arise when analyzing and organizing data in a high-dimensional space. For example, in a pan-European network, the GEANT network\nop{ \cite{Geant2}}, it was recorded (even after traffic sampling) that there were around $10^{9}$ flows distributed over $2^{16}$ ports and $2^{32}$ IP addresses over a 15-minute time interval on a link. They are formidable numbers for analysis in traffic anomaly detection. In addition, in order to identify the possible root cause of traffic anomaly on a time interval, correlating analysis on different traffic features, e.g. source Autonomous System (AS) and destination port number, is often necessary. This implies that the analysis will have to even deal with those numbers in a combinatorial manner, which further complicates the analysis making it hardly implementable. 

In the literature, an extensive body of prior work for traffic anomaly detection exists (e.g. see \cite{guha},\cite{lakhina04},\cite{silvera10},\cite{abdelkefi10}). In these works, unusual abrupt variations in traffic time series, defined as traffic anomalies, raise alarms. Unfortunately, the practical usefulness of the reported alarms is often limited \cite{fontugne10}, \cite{silvera10}, mainly due to the tremendous amount of time and effort additionally required to analyze the root cause of the reported alarms \cite{silvera10}. This results in a pressing need for approaches that perform anomaly detection and root cause analysis jointly.  

To address the above problems, we propose a novel approach, called SENATUS, in this paper. It conducts root-cause analysis jointly with traffic anomaly detection through traffic aggregation and lossy compression. Specifically, SENATUS detects time intervals where anomalous events are suspicious to occur, identifies suspicious aggregate flows, and diagnoses the type of anomaly. The targeted anomaly types are DoS/DDoS attacks and network scans or simply scans.

In brief, SENATUS operates in three stage: {\em election}, {\em voting}, and {\em decision}. Inspired from the concept of a senate, a key idea and the starting point of the proposed approach is to choose or elect a small number of traffic flow sets (termed as senator flows) to represent the total (usually huge) set of traffic flows. Then, on the elected senator flows, traffic anomaly detection is applied, whose results for a time bin together decide or vote if the time bin may be considered as anomalous. Finally, root cause analysis is conducted on each anomalous time bin to identify the root cause of anomaly type for that time bin. The performance of SENATUS is evaluated using traffic traces collected from GEANT, and compared with another histogram-based approach\nop{ (called Apriori in this paper)}. 

The techniques proposed to use in the three stages of SENATUS are, respectively, $K$-sparse approximation of traffic histogram \cite{atef12},  Principal Component Pursuit (PCP) \cite{candes11}, and random tree (RT) machine learning classification \cite{witten11}. These techniques, together with the heuristics proposed in the paper for using them, form the novelty of SENATUS. 
The specific contributions are as follows: 
{\em (i)}  Instead of performing analysis directly on all flows whose original traffic histograms suffer from the curse of dimensionality, we propose to use $K$-sparse approximation to select only the flow sets whose feature values are among the top $K$ on the traffic histogram. This forms the core of the SENATUS election stage to choose senator flows.  
{\em (i)}  SENATUS performs PCP analysis on the time series of each senator flow to detect time bins with abrupt changes, at the voting stage. In addition, the detected time bins with abrupt changes of all senator flows are correlated to flag the most possible anomalous bins. 
{\em (iii)} SENATUS, at the decision stage, performs root-cause analysis for each anomalous time bin, based on the joint application of two intuitive heuristics and a linear time signature and machine learning-based technique. 


The rest is structured as follows. Sec. \ref{sec-2} provides background basics and Sec. \ref{sec-3} gives a detailed introduction of SENATUS. Sec.s \ref{sec-4} and \ref{sec-5} provide evaluation methodology and results. Sec. \ref{sec-6} discusses the related work. Finally, the paper is concluded in Sec. \ref{sec-7}.

\section{Basics of SENATUS} \label{sec-2}
In this section, we describe traffic anomalies and various techniques used in this paper.
\subsection{Targeted Anomalies} 

Table \ref{anomalies charac} lists the categories of anomalies considered in this paper and their traffic characteristics. 

\begin{table} [h!]
\renewcommand{\arraystretch}{1}
\caption{Targeted Anomaly Types}
\label{anomalies charac}
\centering
\begin{tabular}{|c|p{6cm}|}
\hline
Anomaly Type & Traffic Characteristics \\ 
\hline
\hline
DoS & Small or large sized flows sent from one source AS (Autonomous System) via one or multiple source ports to one destination AS on one or multiple destination ports \\
\hline
DDoS & Many small or large sized flows sent from one or many source AS via one or multiple source ports to one destination AS on one or multiple destination ports\\
\hline
Network Scan & \em{Many small sized flows sent from one source AS via one source port to one or many destination AS on one destination port}\\ 
\hline
\end{tabular}
\end{table}

Table \ref{anomalies charac} implies that anomalies of DoS, DDoS or network scan type are often carried by \emph{small size flows}. However, this may lead to misdetection of anomalies that may be present in large size flows. We leave this as future work and highlight that, focusing on small size flows can reduce the risk of false positives in anomaly detection, because large size flows frequently involve benign activities such as bandwidth tests, large file transfers, and high-volume streaming activities \cite{molina12}. 

\subsection{Traffic Histogram}
The proposed SENATUS approach is a traffic histogram-based approach. A {\em traffic histogram} is the distribution of the amount of traffic (in number of flows, packets or bytes) over all possible values of a traffic {\em feature}. A feature is a field in the header of a packet, such as source port, or a \emph{function} of some header field values, such as AS numbers \cite{kind09}. 

While there are many features that may be analyzed, we focus in this paper on four features that are source AS ({\em srcAS}), destination AS ({\em dstAS}), source port ({\em srcPort}) and destination port ({\em dstPort}). This is motivated by traffic characteristics of the targeted anomaly types as discussed in Table \ref{anomalies charac}. 

{\em K-sparse approximation} is a technique proposed for traffic histogram compression \cite{atef12}. It relies on the fact that a traffic histogram may be highly \emph{compressible} if it exhibits a \emph{power-law} decay when sorted, and consequently, one may use the top $K$-feature values to approximate the original traffic histogram. 

More formally, consider a traffic histogram $X$ with $n$ possible distinct feature values (e.g. port 1, port 2, \dots, port $n$). Let $X' \equiv ({x'}_{(1)}, {x'}_{(2)},...{x'}_{(n)})$ denote the sorted histogram, where the coefficients are in the non-increasing order, i.e. ${x'}_{(1)}\geq {x'}_{(2)}\geq...{x'}_{(n)}$. Suppose that the sorted histogram decays according to a power law as, for all $i = 1, 2, \dots, n$,
\begin{equation}
{x'}_{(i)} \leq R \cdot i^{(-\frac{1}{p})}
\end{equation}
where $R$ is a normalization constant and $0 < p \leq 1$ is a scaling parameter. Then, $X$ can be approximated by the first few \emph{``top"-$K$} coefficients, i.e. ${x'}_{(1)}, \dots, {x'}_{(K)}$, with approximation error $\sigma_K$ upper-bounded by \cite{atef12}:
\begin{equation}\label{eq1}
\sigma_K={||X'-X'_K ||}_{2} \leq (ps)^{-\frac{1}{2}}\cdot R \cdot K^{(-s)}
\end{equation}
where $X'_{K}$ has in total $n$ elements defined as $$X'_{K} \equiv ({x'}_{(1)}, \dots, {x'}_{(K)}, 0, \dots, 0)$$ and $s = \frac{1}{p}-\frac{1}{2}$. If the decay of the coefficients is rapid, a small value of $K (<< n)$ can lead to close approximation.

\subsection{Principle Component Pursuit (PCP) }
Principle Component Analysis (PCA) is a statistical tool for high-dimensional data analysis and dimensionality reduction. It basically assumes that the data approximately lie on a low-dimensional linear subspace. Let ${X} \in \mathbb{R}^{n_1 \times n_2}$ be a matrix of interest. The essential idea is to decompose ${X}$ into two components, normal component ${N}$ and anomalous component ${A}$, i.e. ${X} = {N} + {A}$. In the decomposition, the PCA technique attempts to find the matrix ${A}$ such that the matrix ${N}={X}-{A}$ has the lowest possible rank.

More formally, the structural analysis tries to solve the following optimization problem:
\begin{equation}\label{eq1a}
\min_{N,A}\|A\|_{0}, \textrm{ subject to } X = N+A \textrm{ and }rank(N)\leq k
\end{equation}
where $rank(N)$ denotes the rank of a matrix N, $\|\ \|_{0}$ denotes the $\ell_{0}$-norm, i.e. the cardinality of the non-zero elements.

This optimization problem is NP-hard \cite{candes11}. Fortunately, based on recent advances in optimization theory, it has been proved that the nuclear norm, i.e, the sum of singular values, exactly recovers the low rank component $N$ \cite{candes11} while the $\ell_{1}$ norm, i.e, the sum of absolute values, recovers component $A$ in terms of sign and support with remarkable robustness to the outliers in comparison to the $\ell_{2}$ norm \cite{donoho06}.

Accordingly, Eq. (\ref{eq1a}) can be solved using a convex optimization problem called \emph{Principal Component Pursuit} \cite{candes11} as:
\begin{equation}\label{eq2a}
\min_{N,A}\ \|N\|_{*}+\lambda\|A\|_{1},\ subject\ to\  X= N+A,
\end{equation}
where $ \|\ \|_{*}$\ denotes the nuclear norm, i.e., the sum of the singular values of the normal matrix, \ $ \|\ \|_{1}$\ denotes the $\ell_{1}$-norm of the anomalous events matrix $A$, and $\lambda > 0$ is a weighting parameter.


 \subsection{Random Decision Tree}
SENATUS uses the random decision tree (RDT) \cite{rt2} algorithm to find the root cause of anomalies. In fact, RDT is an ensemble of decision trees. The process for generating a tree is as follows. First, it starts with a list of features or attributes from the data set. Then, it generates a tree by randomly choosing one of the features without using any training data. The tree stops growing once the height limit is reached. Then, it uses the training data to update the statistics of each node. Note that only the leaf nodes need to record the number of examples of different classes that are classified through the nodes in the tree. The training data is scanned exactly once to update the statistics in multiple random trees. The further explanation of RDT can be found in Section \ref{sec-3C}.
\section{Detailed SENATUS Approach} \label{sec-3}

As discussed earlier, SENATUS can be divided into three stages. In this section, we describe all these stages in detail.

\subsection{Election Stage}

In this stage, senator flows and senator subspace are defined, which are used in further analysis in the later stages.
\begin{table}[h!]
\label{hts}
\caption{Heuristics}
\label{heuristics}
\centering
\begin{tabular}{|c|c|}
\hline
Heuristic & Definition \\
\hline
\hline
H1 & Small packet count per flow: \# of packets $\leq \alpha$\\
\hline
H2 & Small byte count per flow: \# of bytes $\leq \beta$\\
\hline
\end{tabular}
\end{table}

The traffic traces are pre-filtered using two heuristic H1 and H2 defined in Table \ref{heuristics}. In H1, \emph{small size flows} are defined as those flows whose packet counts are not larger than threshold value $\alpha$ and in H2, \emph{small size flows} are defined as those flows whose byte counts are not larger than threshold value $\beta$, in a time bin. A more detailed investigation of the effect of the threshold values will be presented in Section \ref{sec-4}. 

After this filtering, for every measurement time bin, K-sparse approximation is applied to the flow number histogram of each of the four traffic features (srcAS, dstAS, srcPort, dstPort) on the time bin. Here, our approach behind pre-filtering traffic before applying K-sparse approximation is that, with pre-filtering, it is more likely that an anomalous feature value is included in the selected top $K$ components. 

After $K$-sparse approximation is applied for a traffic feature (e.g. srcAS) on a time bin $t$, $K$ top values of this feature $j$ for this time bin are obtained. Let $I_j(t)$ denote the set of these $K$ top values of the feature $j$ on the time bin $t$. Suppose there are $N$ time bins in the traffic trace. Let $I_j$ be the consolidation of these $N$ sets of such feature values, i.e. $I_j = I_j(1) \cup \cdots \cup I_j(N)$.   

Then, for every feature $j \in$ \{{\em srcAS, dstAS, srcPort, dstPort}\}, if it has a value $i$ in $I_j$, i.e. $i \in I_j$, this feature value defines a {\em senator flow} or simply {\em senator} for the feature $j$.  We remark that each senator is indeed a flow aggregate in which all flows have the same feature value. 

{\em The senator subspace} is a three-dimension flow count matrix $Y(t,i_j,j)$ defined on time $t (= 1, \dots, N)$ and feature value $i_j \in I_j$ across all features $j \in$ \{{\em srcAs, dstAs, srcPort, dstPort}\}. Later analysis will be applied on this senator subspace. 
\subsection{Voting Stage}
At this stage, SENATUS analyzes each senator's time series using PCP to detect abrupt changes on the time series. The detected abrupt variations, called votes, are then correlated on time to identify or vote the most likely anomalous time bins. 


For every feature $j \in$ \{{\em srcAS, dstAS, srcPort, dstPort}\}, let $X(t, i_j)$ be its traffic amount time series matrix. Specifically, the element at $(t,i_j)$ of $X$ records the number of flows that have the same feature value $i_j$ (e.g. srcPort 80) at time bin $t$ in the measurement period. Essentially, $X(t, i_j) = Y(t, i_j, j)$ with $j$ fixed to be the considered feature.

Applying the PCP technique described in Section \ref{sec-2}.C to the time series matrix $X$, where $n_1=n$ i.e. the number of time bins in the measurement period and $n_2=I_j$ i.e. the number of senators from feature $j$, the corresponding anomalous events matrix $A$ is obtained. The positive-value elements in the obtained anomalous events matrices are referred to as {\em votes}. 

For any time bin $t$, a feature $j$ (e.g. srcPort) is flagged anomalous if (at least) one of its values in $I_j$ makes a vote on this time bin, or in other words, at least one senator time series of this feature has abrupt variation on $t$. For the time bin $t$, if all features \{{\em srcAS, dstAS, srcPort, dstPort} \} have been flagged anomalous, this time bin is considered to be an anomalous time bin. 


\subsection{Decision Stage} \label{sec-3C}

In this stage, the attempt is made to diagnose the root-cause for each anomalous time bin. In particular, our objective is to investigate if the traffic behavior on an anomalous time bin is due to one of the focused anomaly types listed in Table \ref{anomalies charac}.

\subsubsection{Suspicious Flow Aggregate} 
Let $m_j$ denotes by the cardinality or the number of senator members of $I_j$. In addition, let a flow aggregate is defined by  srcAS, dstAS, srcPort and dstPort. Then, for the time bin, there are $M = m_{srcAS} \times m_{dstAS} \times m_{srcPort} \times m_{dstPort}$ such flow aggregates, which are called suspicious flow aggregates. These combinations essentially tell that we consider flows that might be originated from any suspicious source AS at any suspicious port and target at any suspicious destination AS at any suspicious port. Note that, for some of these combinations, the number of flows in the flow aggregate is zero. Such flow aggregates will be skipped in later analysis. We call the remaining ones the effective suspicious flow aggregates.

\subsubsection{Root Cause Analysis}

After identifying the set of suspicious aggregate flows, we aim to infer the event that has caused the alarm while flagging the time bin as anomalous. Let $\theta_1$, $\theta_2$ and $\theta_3$ denote the three thresholds to be used in the algorithms for classifying if the anomaly type is DDOS, DOS or network scan respectively. 

The root cause analysis algorithm  in SENATUS is a signature-based algorithm. Specifically, for every effective suspicious flow aggregate, it performs the following:

\begin{enumerate}
 \item For every $dstIP$ that is included in the $dstAS$ of the suspicious flow aggregate, find the number of flows that are destined to this dstIP, regardless of their $srcAS$, $srcPort$ or $dstPort$. Take the maximum of all such numbers and call it the anomaly intensity. Compare this intensity with $\theta_1$. If the former is greater, output is DDoS. Otherwise, perform the next. 
 \item For every $\{srcIP, dstIP\}$ pair included in the $\{srcAS, dstAS\}$ pair of the suspicious flow aggregate,
perform similarly as above: Find the number of flows that have the same srcIP and dstIP, regardless of their $srcPort$ or $dstPort$. Take the maximum of all such numbers. Compare this with $\theta_2$. If the former is greater, output is DoS. Otherwise, perform the next.

\item For the $dstPort$ of the suspicious flow aggregate and every $srcIP$ that is included in the $srcAS$ of the aggregate, find the number of flows that are originated from this srcIP and destined to this dstPort, regardless of
their $srcPort$ or $dstAS$. Take the maximum of all such numbers and compare this intensity value with $\theta_3$. If the former is greater, output is Network Scan. Otherwise, repeat these steps for the next effective suspicious flow aggregate.
\end{enumerate}
The above procedure is repeated until an attack signature comparison is successful. Or, in the end, the anomaly
type cannot be identified and in this case the alarm is reported as false positive.

\subsubsection{Threshold Values}

We highlight that the thresholds $(\theta_{i}, i=1, 2, 3)$ are key parameters in the root cause analysis. In particular, the RDT algorithm \cite{rt2} is used to find $(\theta_{i}, i=1, 2, 3)$, which turned out to be eminently suitable since it provides high classification rate in our scenario, fast and easy to interpret, as implied by the optimality of probability estimation by RDT \cite{rt1}. 

In our algorithm, each anomaly is mapped as a point into a space where anomalies are classified based on their intensities. Under this taxonomy, we create a set of labeled instances that consist of the attribute: intensity in number of flows, which is mapped to one of the three anomaly classes: DoS, DDoS and Scans. The labeled instances serve as input to the RDT learning algorithm that outputs a tree which indicates the range of intensities per anomaly class.

Our algorithm works as follows. It is an iterative algorithm. For time $T = 1, 2, \dots, N$, the inputs are the set of unknown anomalies at this time and the set of previously labeled anomalies for times $[1,T-1]$. The algorithm first applies the decision tree technique on the labeled items of anomalies and their corresponding intensity. The output is a tree $DT_{[1,T-1]}=(Br(i,j),Class(j)), i\leq size(DT), j \leq size(Br)$ where each path constitutes a set of branches from the root to a leaf. A branch $j$ of a path $i$, $Br(i,j)$, introduces an upper or a lower bound of an anomaly intensity while a leaf of a path $i$: Class(i) corresponds to a class of anomaly. 

We then explore the output tree and map it into a set of association rules to enable classification of anomalies based on their intensity. A rule is an antecedent $\{Br(1,i),...Br(n,i)\}$ which represents the $i_{th}$ path of the tree and a consequent, i.e, a class of anomalies. Since only one attribute (anomaly intensity) is adopted in the learning process, the branch $Br(n,i)$ from each leaf to its direct parent defines each association rule antecedent which is, defined as comparator, i.e, ${\geq,\leq}$ and a value, i.e, a threshold of anomaly intensity. The threshold values are then extracted by simple parsing of the set of rule antecedents: rule antecedents which introduces an upper bound of intensity for a class of anomaly are ignored, while those which introduce a lower bound (comparator=$\{\geq\}$) are considered. A lower bound of anomaly intensity in association rule antecedent represents a candidate threshold. The output threshold $\theta_{i}, i=1..3$ value for a given class of anomalies is the minimum among all candidate thresholds.


\section{Evaluation Methodology}\label{sec-4}
This section is divided into three parts.  In the first part, real data sets (chosen for our evaluation) are described. In the second, ground truth data construction is explained and finally, analysis on chosen parameters of SENATUS is provided.


\subsection{Dataset} 
The measurement dataset used in this paper is comprised of traffic traces collected from the GEANT network\nop{ \cite{Geant2}}. GEANT is a pan-Europe backbone network interconnecting European NRENs (National Research and Educational Networks) and provides them access to other NRENs and the Internet using dedicated links. The traffic traces in the adopted dataset were collected from the following four links: (1) a peering link between the Internet and the Frankfurt router in GEANT (Trace $\mathcal{A}$); (2) a peering link between the Internet and the Vienna router in GEANT (Trace $\mathcal{B}$); (3) a peering link between the Internet and the Amsterdam router in GEANT (Trace $\mathcal{C}$); (4) a peering link between the Internet and the Copenhagen router in GEANT (Trace $\mathcal{D}$).

The four traces\footnote{They are available from GEANT on request\nop{ \cite{Geant2}}.} were collected during a 18-day measurement period in June -- July 2011, and involve flow records over 15-minute measurement time bins at a sampling rate of $1/1000$. A flow record involves different information fields such as source and destination IP address and AS numbers, source and destination ports, transport protocol (TCP/UDP), the duration of a flow (in second) and the flow size in packets and bytes. We analyzed all the four traces and found that that they have low rank traffic metrics and sparse abrupt variations \cite{lakhinaimc04}. 

\subsection{Ground-Truth Construction}
Without the ground-truth data, the process of manually inspecting anomalous time intervals (each contains hundreds or thousands of anomalous flows) is an onerous process. To make the construction of ground-truth feasible and tractable, we adopt a combined method. Specifically, we run both SENATUS (using a given heuristic) and the anomaly detection approach
that SENATUS will be compared with on the dataset traces. For each time bin, if both approaches flag the same anomaly
type, then the anomaly is added to the ground-truth. However, if for a particular time bin, only one of them flags an anomaly or each flags a different type of anomaly, we extract the following four tuple features: srcIP, dstIP, srcPort and dstPort for each of the flagged anomalous flows and do manual inspection in the following way. We draw a scatter plot per each couple of traffic features in addition to a graphlet of communication pattern \cite{fontugne2010} and check whether the label suggested by either method matches visual inspection of the graphs. If there is match, the alarm is added to the ground-truth; otherwise, the alarm is considered as a false positive. 

\subsection{SENATUS Parameters} 

We discuss in this subsection the impact of tuning parameters on the performance of SENATUS. The study of this impact is based on evaluation of SENATUS' output with the tuning of its various parameters, predicated on the ``ground-truth" constructed in the previous section. The various parameters associated with SENATUS, and their constraints, are summarized in Table \ref{Senatusparam}.

\begin{table}[ht]
\caption{SENATUS Parameters}
\label{Senatusparam}
\centering
\begin{tabular}{|c|c|c|}
\hline
Parameter & Description &Constraints \\
\hline
\hline
$\alpha$ (for H1) & flow size in packets &small\\
\hline
$\beta$ (for H2) & flow size in bytes &small\\
\hline
$K$ & number of senator &small\\
\hline
$\lambda$& PCP weighting parameter &$\geq 2$\\
\hline
$j$& flow aggregation level &$[1,4]$\\
\hline
\end{tabular}
\end{table}


\subsubsection{Traffic Filtering Heuristics H1 and H2}

As previously discussed, traffic pre-filtering tends to concentrate anomalies in the top-$K$ feature values, thus reducing the number of components required for traffic histogram approximation. In this subsection we study the range of both parameters $\alpha$ and $\beta$ and explain our choice of their values.
\begin{figure}[thb!]
\centering
\subfigure[]{
   \includegraphics[width=0.228\textwidth]{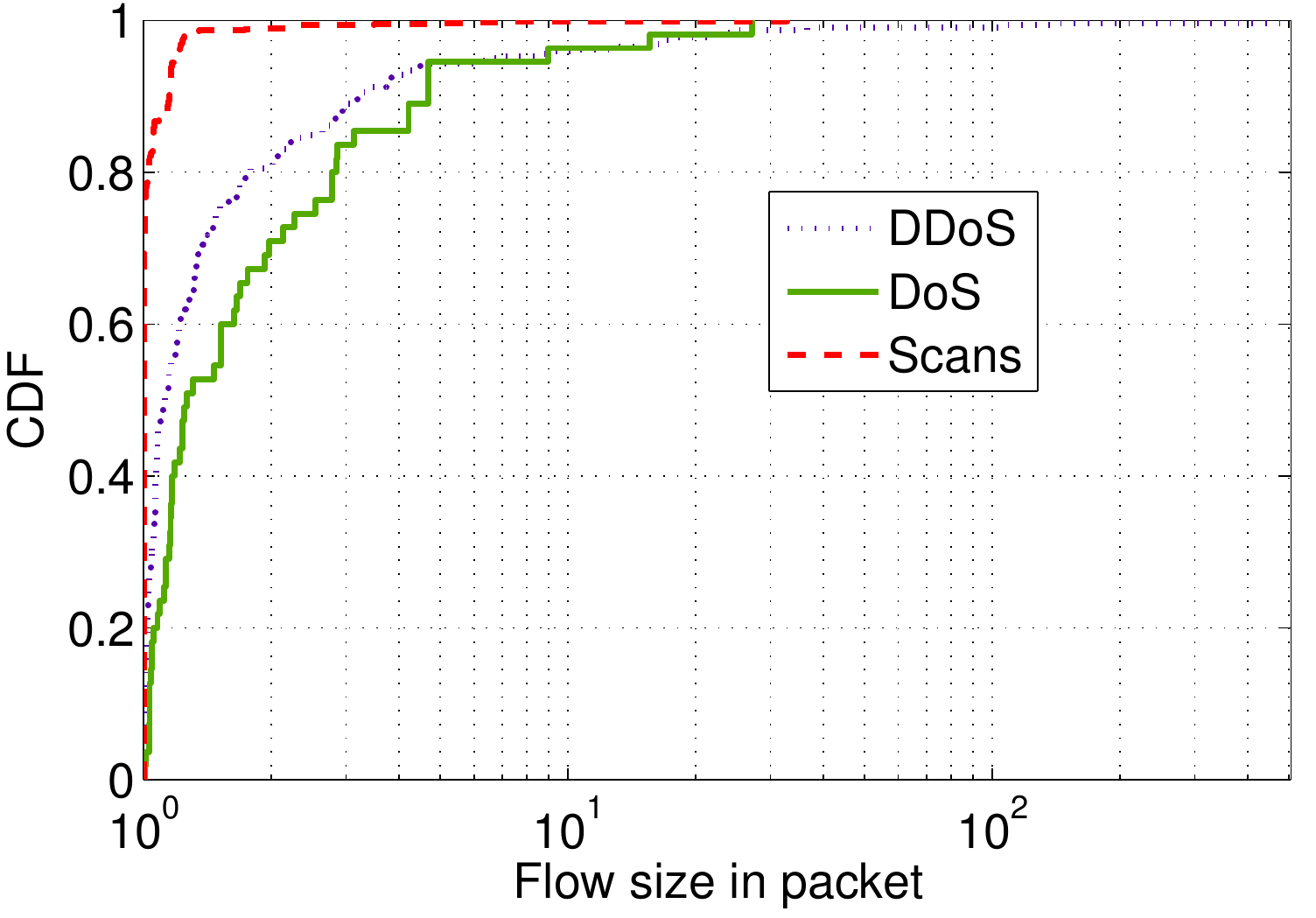}
 }
 \subfigure[]{
  \includegraphics[width=0.212\textwidth]{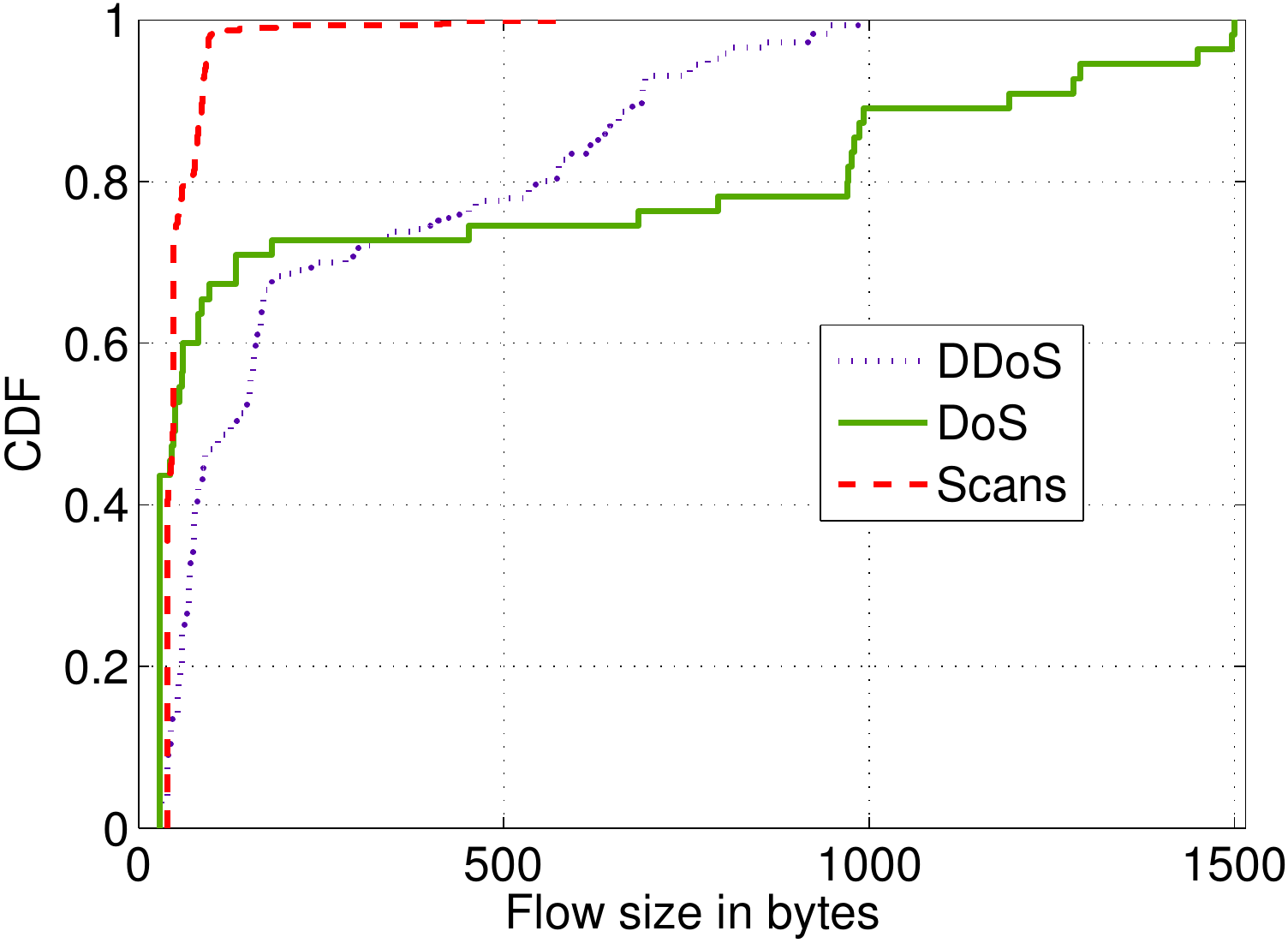}
 }
 \vspace{-0.25cm}
\caption{Average flow size per anomaly distribution}\label{anomdistributions}
 \vspace{-0.5cm}
\end{figure}

Related to heuristic H1, previous study (e.g. \cite{xu05}) has reported that most of the encountered anomalies (including scans) in datasets are carried by flows with number of packets in the range of $[1,3]$. Authors of \cite{gates06} also claim that most of the detected scans are carried by flows having a packet count number $\leq 2$. 

Our investigation of the flow size in the constructed ground-truth is shown in Figure \ref{anomdistributions} (a). The figure illustrates the distribution of the average flow size in term of packets for the inspected anomalies in our traces. Figure \ref{anomdistributions} (a) depicts that while most of the anomalous flows have an average size of one packet, anomalies involving flow-size of less than or equal 3 packet are in the order of 85\% of all attacks in our data set.

Related to heuristic H2, several previous studies have tried to investigate the validity of using H2 as a heuristic to filter out traffic carrying network abuse attacks. Authors of \cite{xu05} show that most of the detected anomalies (including scans, worms, etc.) in their dataset are carried by small flows having a byte count $\in [40,144]$. Authors of \cite{liu11} give narrower range of byte count and show that over $99\%$ of the detected DDoS attacks in CAIDA traces have a packet size falling in the range between $40$ and $60$ bytes.

Our inspection of the attacks in the constructed ground-truth shows similar results. The average flow size distribution in terms of number of bytes is illustrated in Figure \ref{anomdistributions} (b). Although of a long tail due to variable size of the flagged anomalous flows, most frequent DoS/DDoS and scans in our traces are of a small size ($\leq 64\ bytes$). For example 52\% of the detected DOS, while 99\% of the detected scans carry flows of size 60 bytes.

In the remaining, the threshold values $\alpha$ and $\beta$ used in our framework are set to respectively $3$ and $64$.

\subsubsection{The choice of K} 

 We choose $K$ such that it realizes an average approximation error $\sigma_{K} \in [0.01, 0.3]$ depending on the measurement trace and the type of traffic under analysis. We assume that the resultant $K$ value under such an approximation error is an acceptable ``information-loss'' tradeoff.

\begin{figure}[thb!]
\centering
\subfigure{
   \includegraphics[width=0.22\textwidth]{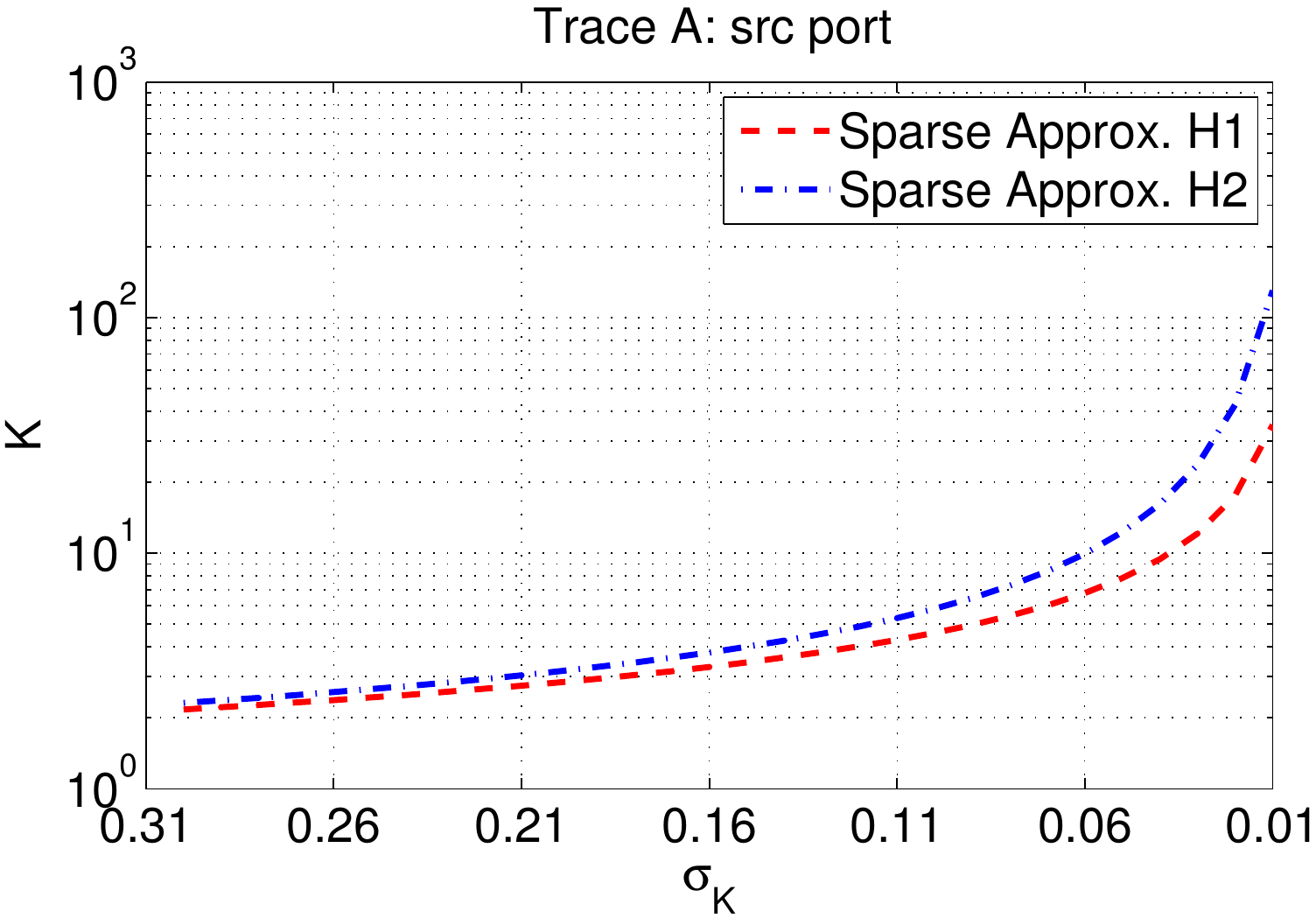}
 }
 \subfigure{
  \includegraphics[width=0.22\textwidth]{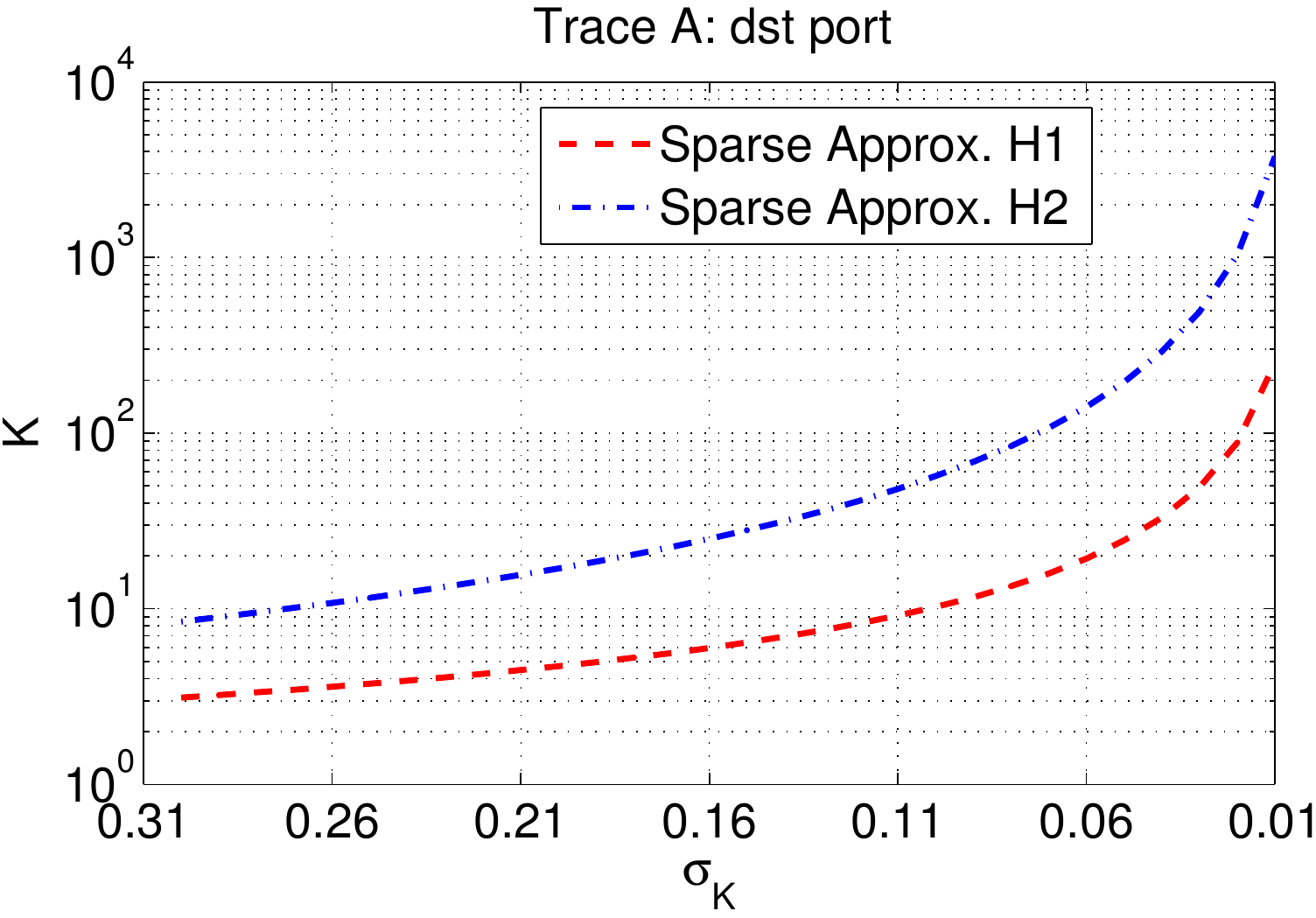}
 }
\subfigure{
   \includegraphics[width=0.22\textwidth]{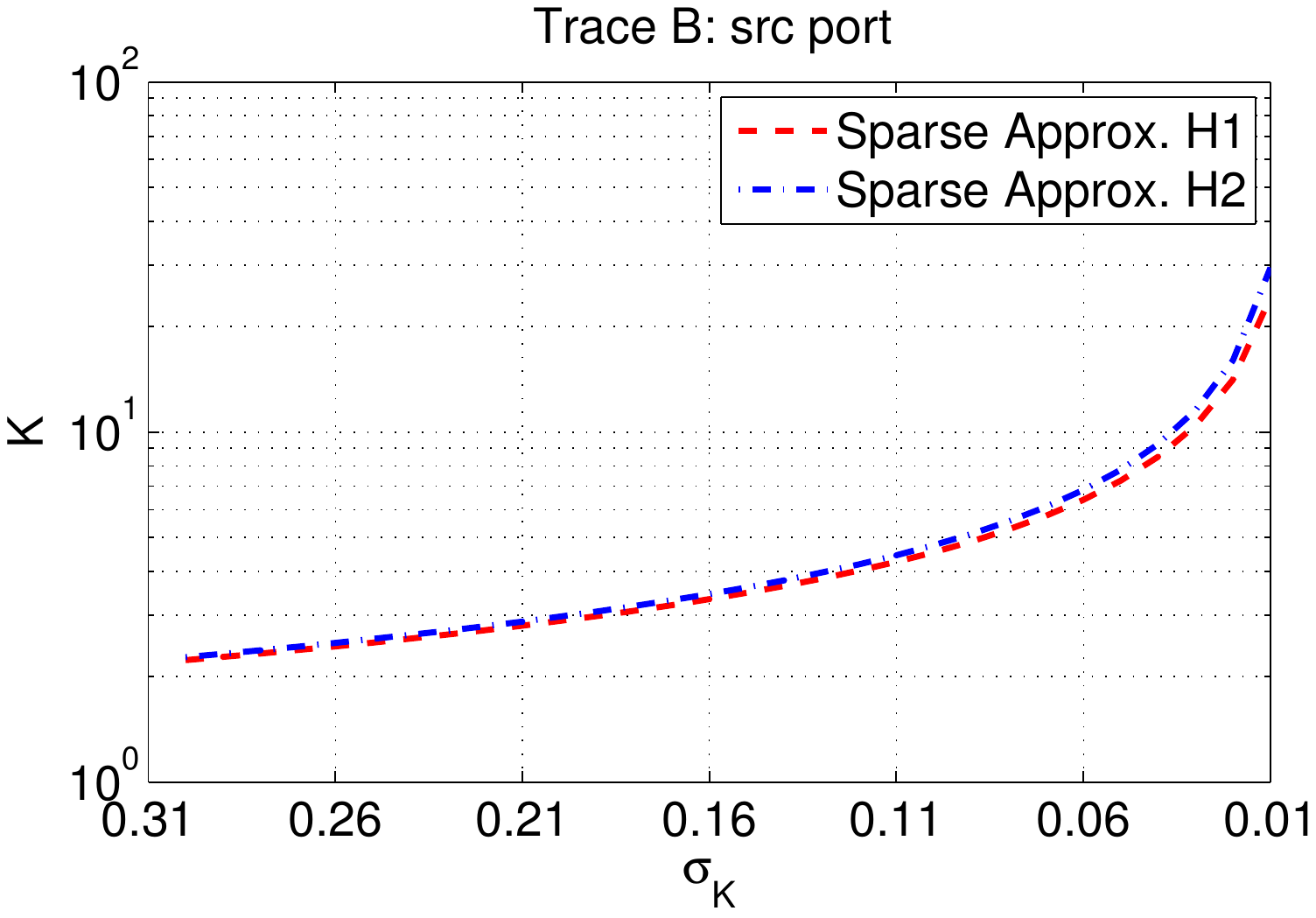}
 }
 \subfigure{
  \includegraphics[width=0.22\textwidth]{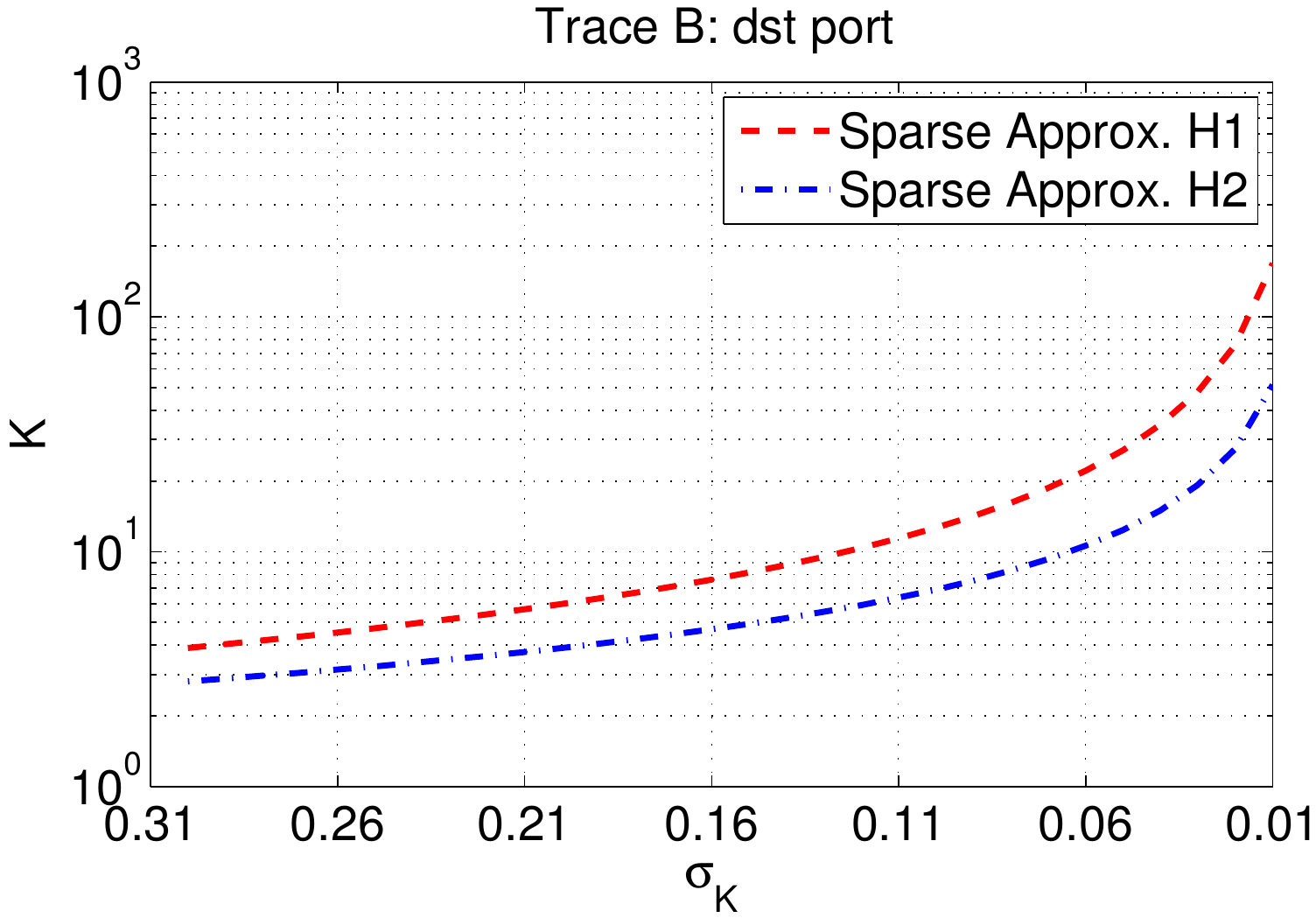}
 }
\subfigure{
   \includegraphics[width=0.22\textwidth]{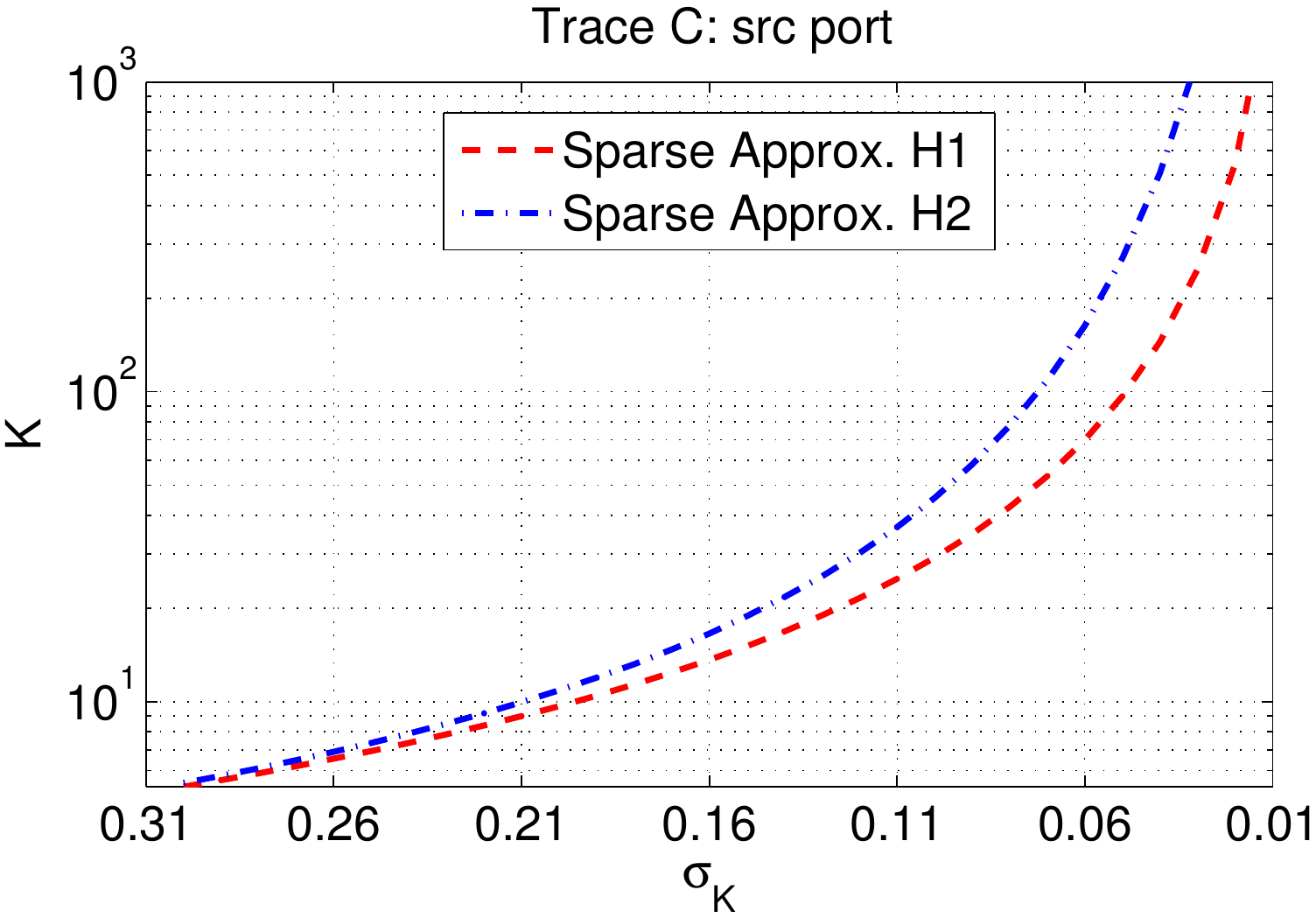}
 }
 \subfigure{
  \includegraphics[width=0.22\textwidth]{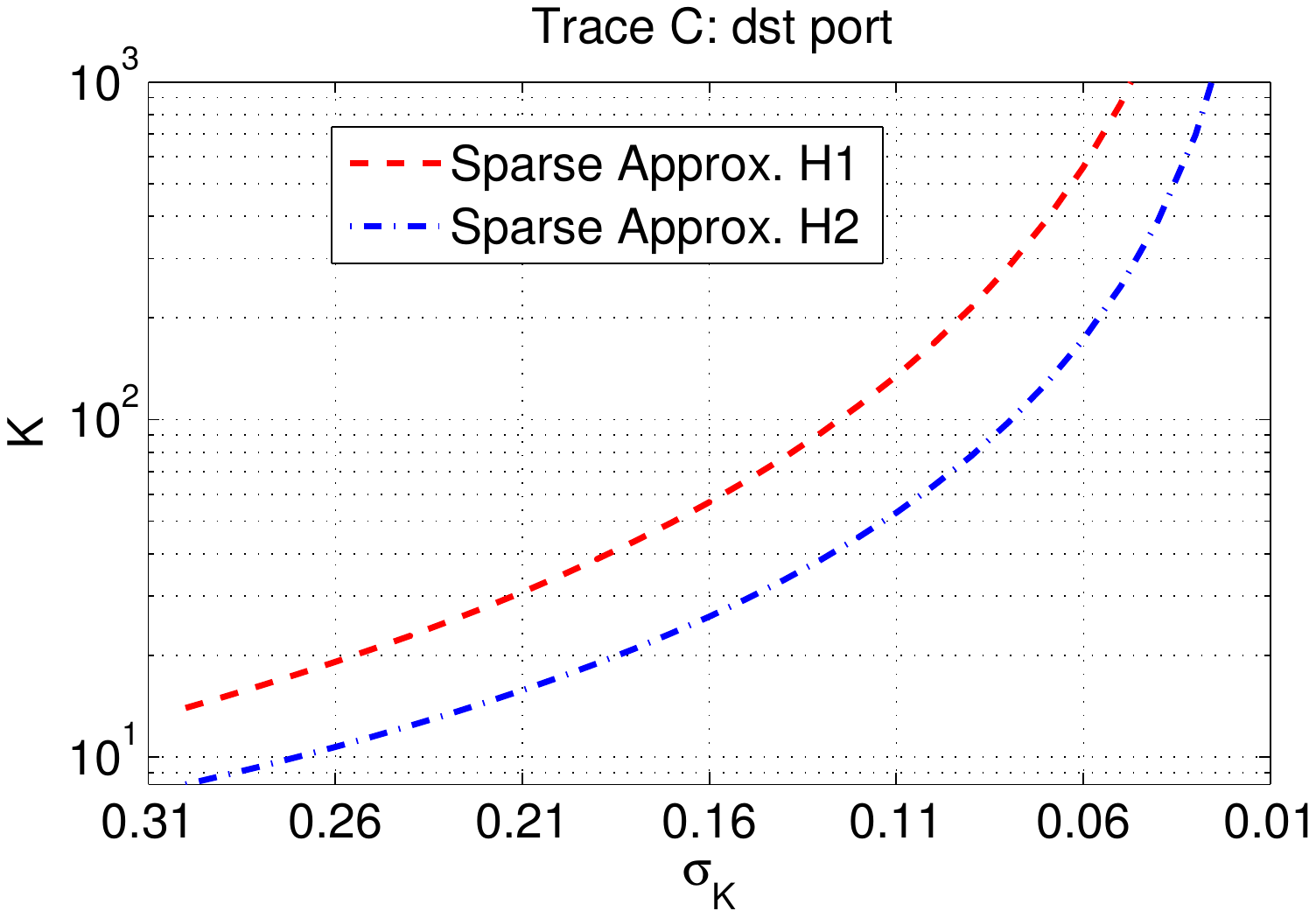}
 }
 \subfigure{
   \includegraphics[width=0.22\textwidth]{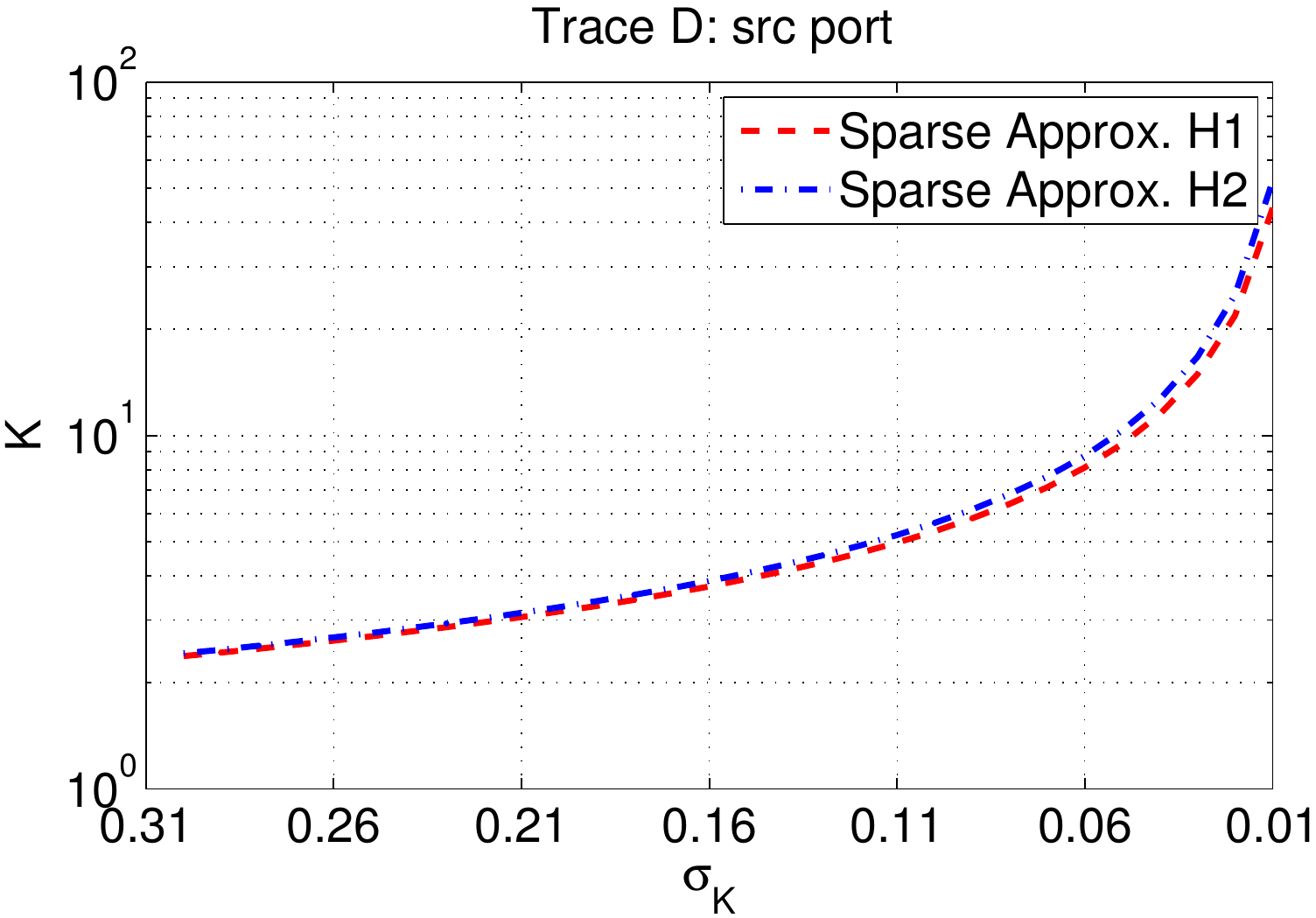}
 }
 \subfigure{
  \includegraphics[width=0.22\textwidth]{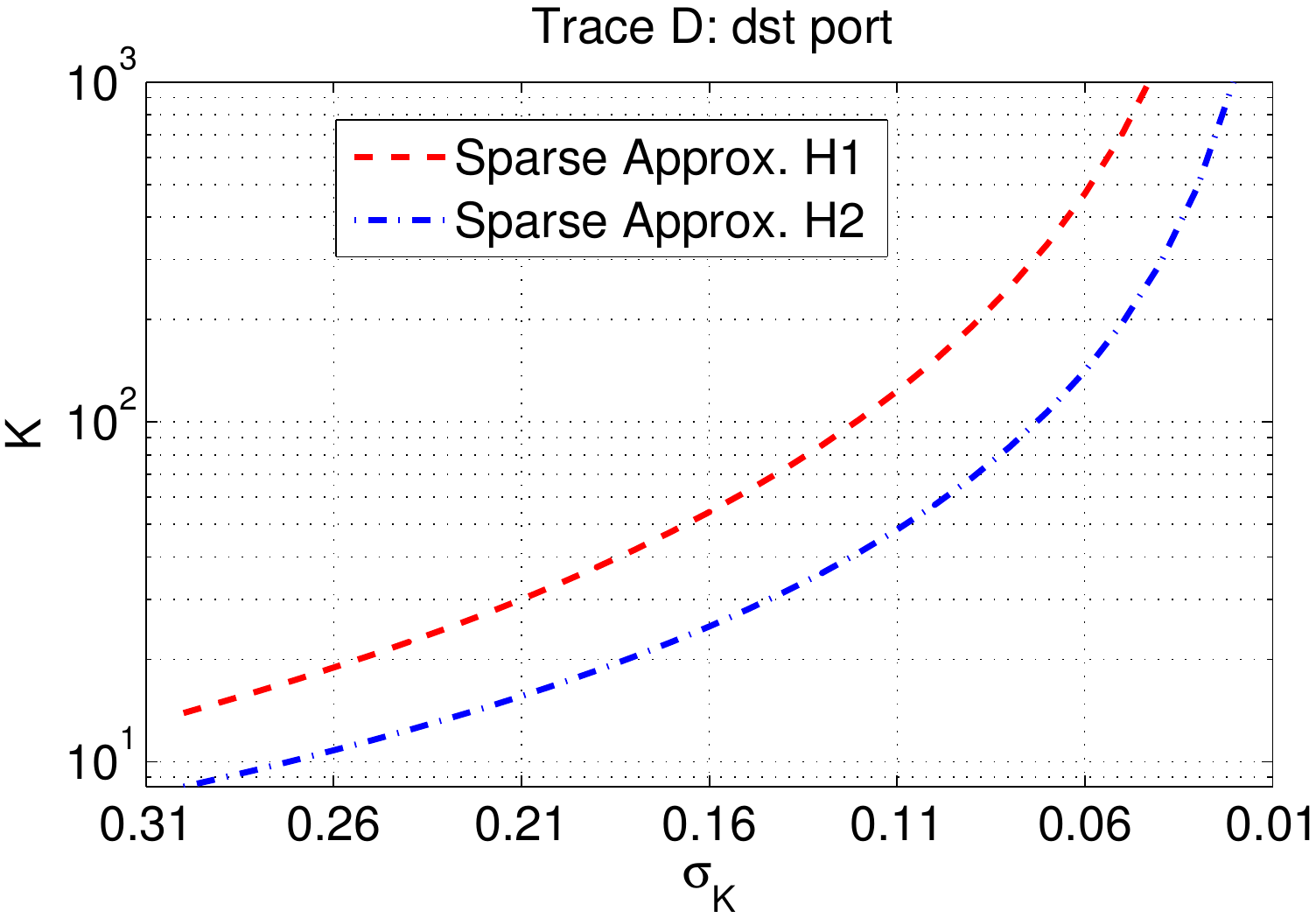}
 }
\caption{$K$ value as a function of the approximation error}\label{kerror}
\vspace{-2mm}
\end{figure}

Figure \ref{kerror} illustrates the range of the $K$ value which achieves an average approximation error in the range [0.01, 0.3] per pre-filtering heuristic, for each of the\nop{ collected} traces\nop{ during the measurement interval}. Expectedly, as the approximation error decreases the required number of coefficients for traffic histogram approximation exponentially increases. The figure additionally reports that the value $K$ can vary from several tens to hundreds in order to achieve a targeted approximation error, depending on the measurement trace and the pre-filtering heuristic.

\begin{table}[th!]
\caption{Approximation errors under $K=20$}
\label{k}
\centering
\begin{tabular}{|c|c|c|c|c|c|}
\hline
Feature &Heuristic  &$\mathcal{A}$ &$\mathcal{B}$ &$\mathcal{C}$ &$\mathcal{D}$ \\
\hline
\hline
\multirow{2}{*}{Srcport} & H1 &0.03 & 0.02 &  0.14 & 0.02\\
\cline{2-6}
& H2& 0.02 & 0.01 &  0.12 & 0.02\\
\hline
\multirow{2}{*}{Dstport} & H1 & 0.18 & 0.03 & 0.18 & 0.18\\
\cline{2-6}
& H2& 0.06& 0.06& 0.25 & 0.25\\
\hline
\end{tabular}
\end{table}

To avoid complex tuning of the $K$ parameter and motivated with the observation that the value of $K$ is stable over time \cite{atef12}, we choose for simplicity one value of $K$, i.e. $K=20$, for all traces under both heuristics. Table \ref{k} illustrates the resultant average approximation error for the four measurement traces using each of the proposed heuristics under $K=20$.

\subsubsection{The choice of the PCP Tuning Parameter}

Since PCP aims to minimize the weighted combination of the nuclear norm and the $\ell_{1}$-norm, one has to identify an appropriate value of the weight parameter $\lambda$ such that the matrix $A$ captures the maximum number of anomalies with the least false positive rate. In PCP, parameter $\lambda$ is also expressed as\nop{in the following form} \cite{candes11}:
\begin{equation}
\lambda= \frac{C}{\sqrt{\max \{N,K \}}}, C \in \mathbb{R}
\end{equation}
where $N$ and $K$ are the dimension of the senators' subspace.
\begin{figure}[th!]
\centering
\subfigure{
   \includegraphics[width=0.22\textwidth]{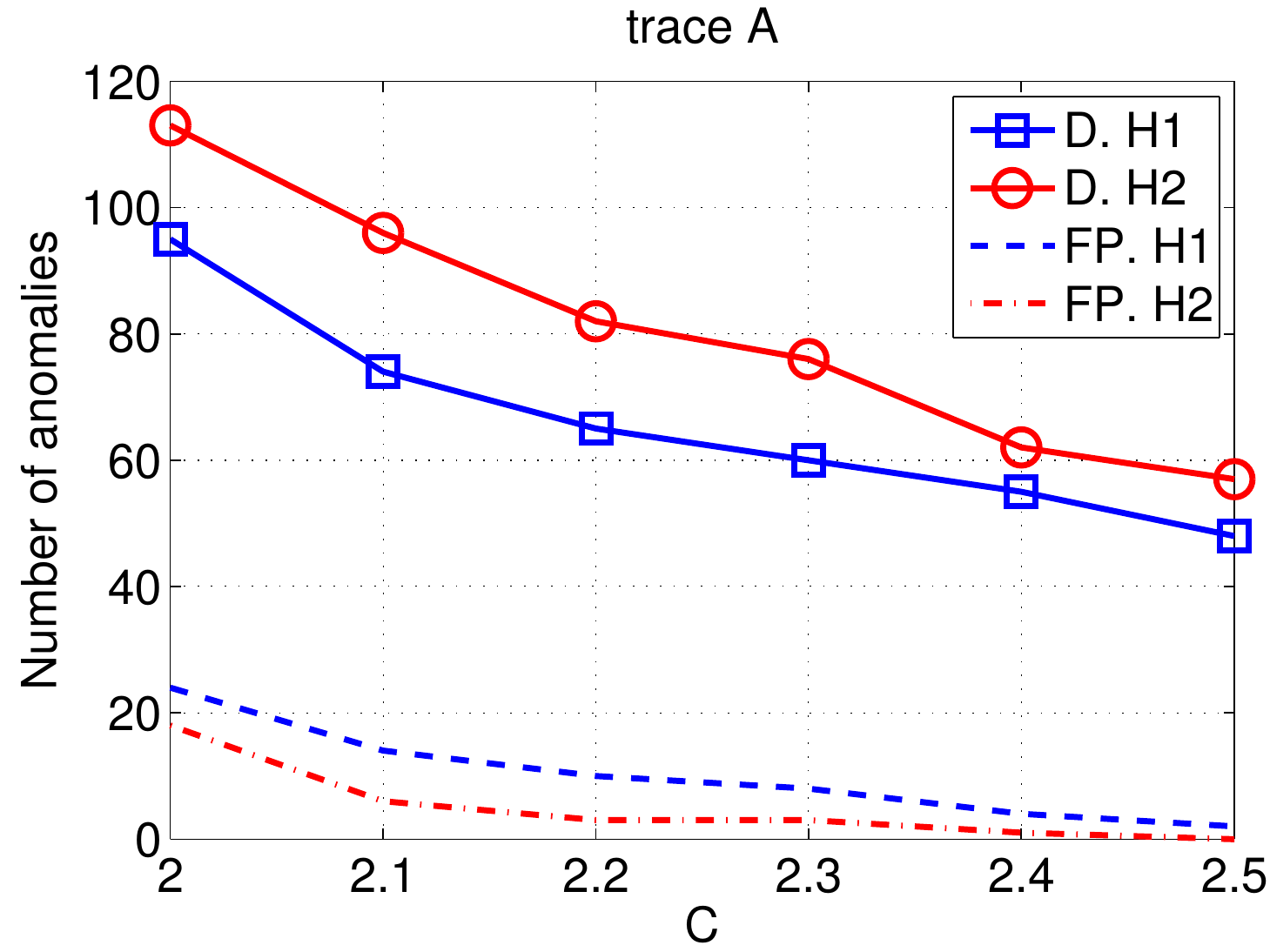}
}
\subfigure{
   \includegraphics[width=0.22\textwidth]{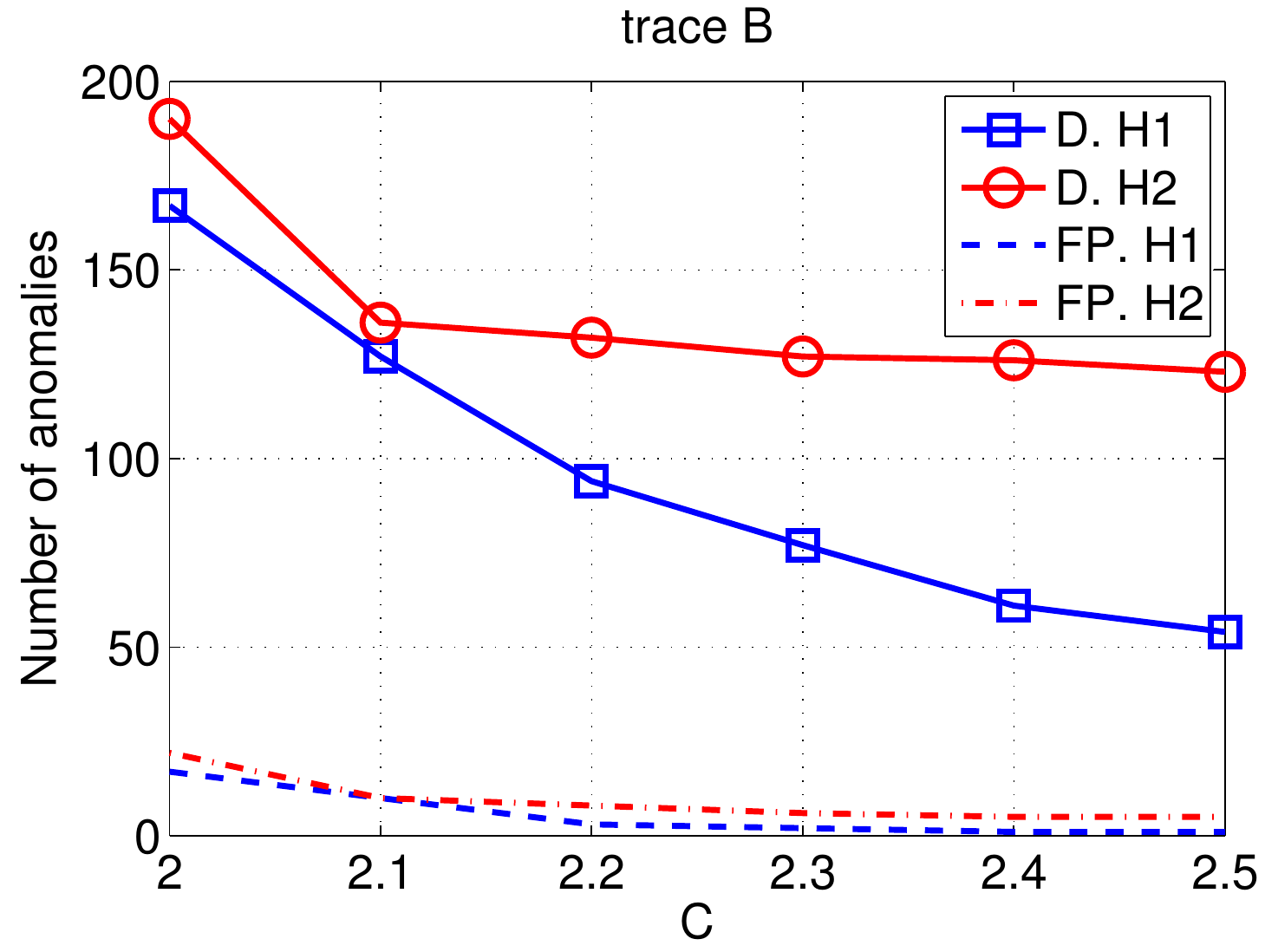}
}
\subfigure{
   \includegraphics[width=0.22\textwidth]{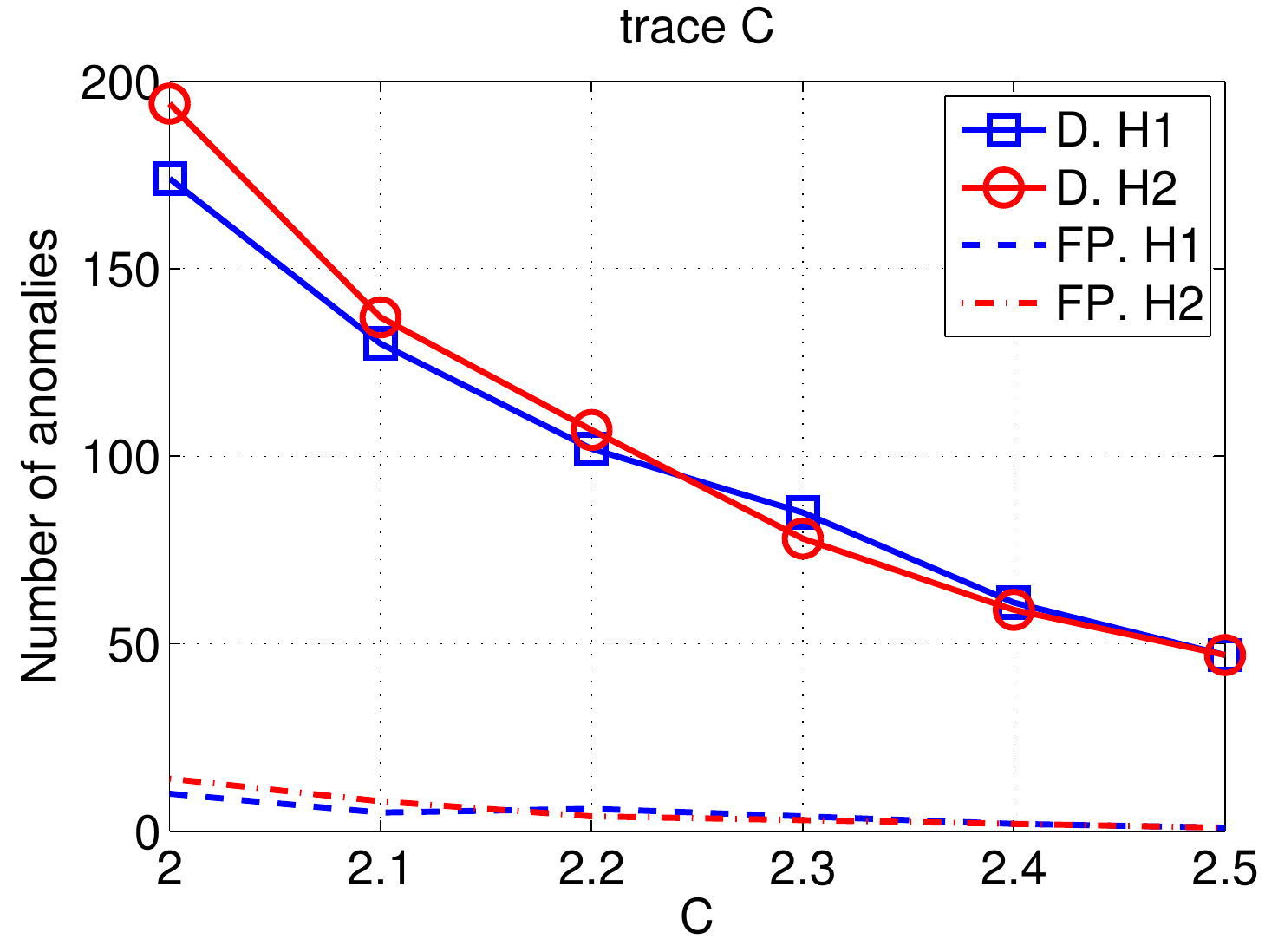}
}
\subfigure{
  \includegraphics[width=0.22\textwidth]{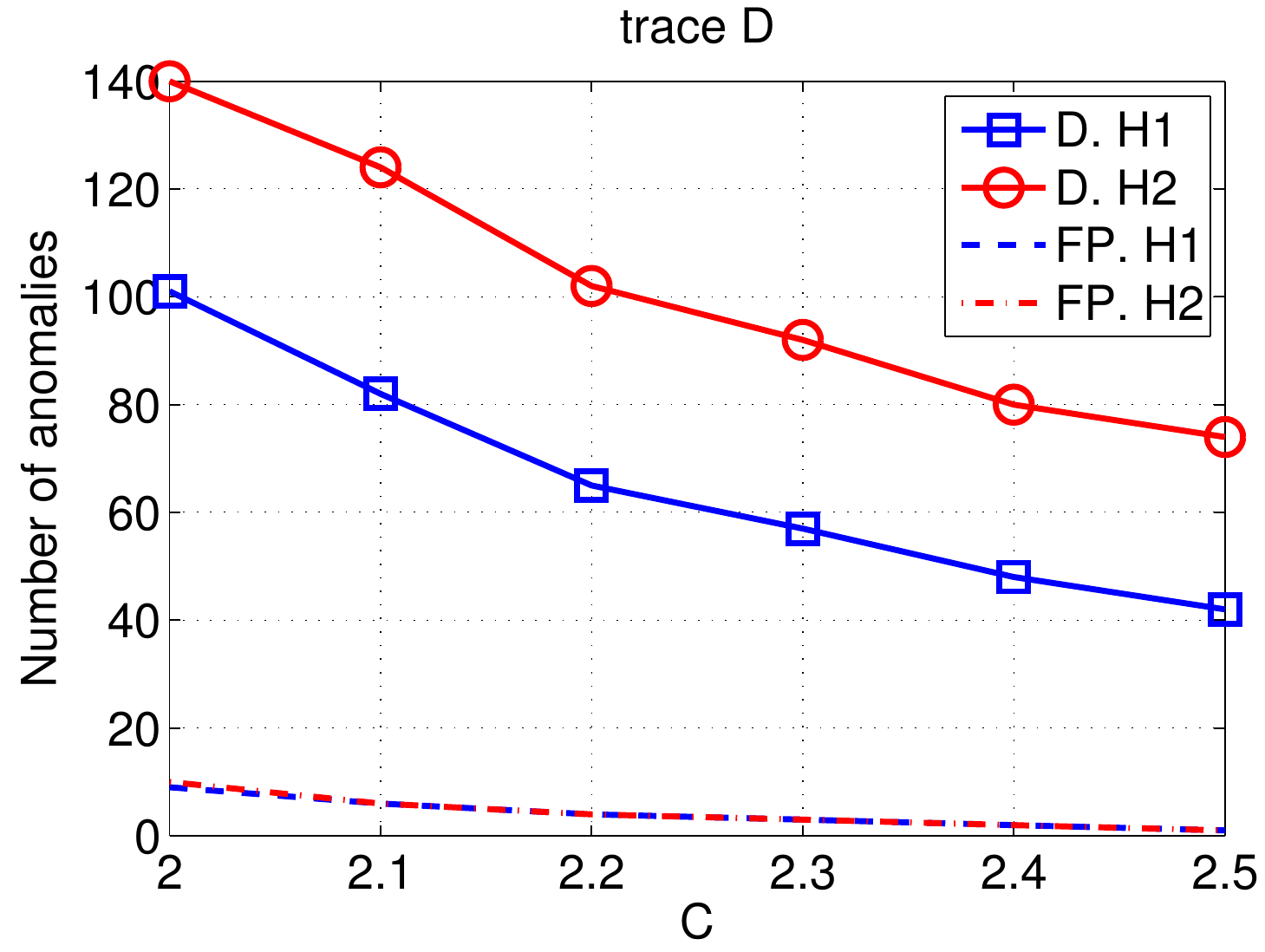}
}
\caption{Detection and false positive rates as functions of $C$}\label{dfplambda}
\vspace{-0.3cm}
\end{figure}

It has been previously shown that $C=2$ is appropriate for anomaly detection in traffic time series \cite{abdelkefi10}. We base our analysis on the previous observation and tune the parameter $\lambda$ to find an ``optimal" value that achieves the ``best" detection-false positive tradeoff. To illustrate this\nop{ tradeoff}, Figure \ref{dfplambda} is presented\nop{ which shows}showing the number of detected anomalies and the false positives as functions of the parameter $C$. The figure shows that both the detection and false positive rates decrease as the value of $C$ increases. For example, when H2 is used, 113 anomalies are detected with 9 false positives for the value of $C=2$ while only 57 anomalies are detected with 1 false positives for $C=2.5$, both in trace $\mathcal{A}$. The figure additionally shows that while the number of false positives, when H1 is chosen, is higher than those when H2 is used, it remains low for all values of $C$. In the remaining of the evaluation we choose the value $C=2$ for both heuristics H1 and H2.

\section{Results} \label{sec-5}
In the first part, the Apriori approach is described with which SENATUS is compared. Then, the results are provided.
\subsection{Apriori Approach}
This approach is based on \cite{brauckoff09imc}. Specifically, it first uses histogram-based detectors to identify suspicious flows and then applies association rule mining to find and summarize anomalous event flows.
For the former, it uses Kullback-Leibler (KL) distance, and for the latter, it makes use of the Apriori algorithm introduced in \cite{agrawal94}. 

The KL distance idea has been widely applied for anomaly detection in previous works \cite{kind09, ramah09, brauckoff09imc}. Motivated with the basic assumption that anomalies deteriorate traffic histograms, the KL distance identifies anomalous time intervals by measuring the similarity between the current traffic histogram and a reference histogram. More formally, given a discrete distribution $q$ and a reference distribution $p$, KL distance $D$ is defined as follows:
\begin{equation}
D(p||q)=\sum_{i}^m p_{i} log(p_{i}/q_{i}).
\end{equation}
To compare with SENATUS, we apply the KL distance on random projections (hash functions) of traffic histograms at each of the 4-tuple features on each of the measurement intervals. Particularly, the hash function randomly places each traffic feature value into a set of lower-dimensional bins, which represents a \emph{lossless compression} process. In addition, the distribution from the previous time interval is used as the reference distribution $p$ \cite{brauckoff09imc}. The KL distance value which exceeds a predefined threshold for any time interval $t$ serves to detect anomalies.

After the set of anomalous time bins are identified by the KL distance, root-cause analysis is performed, extracting the set of candidate anomalous flows responsible for the flagged anomalies using a flow pre-filtering algorithm \cite{brauckoff09imc}. This algorithm generates the meta-data that is suspicious to contain the highest amount of anomalous flows. Such flows are further extracted using a frequent item-set mining algorithm (Apriori) proposed in \cite{agrawal94}. 

\subsection{Detected Anomalies per Type}

Table \ref{anomaliespertype} presents the number of anomalies per type found by each method. The table shows that SENATUS generally detects more anomalies (particularly network scans) than Apriori. However, Apriori detects more DDoS attacks for trace $\mathcal{C}$ and $\mathcal{D}$. We also observe that while SENATUS using H1 or H2 detects more DoS attacks for traces $\mathcal{A}$,$\mathcal{B}$ and $\mathcal{D}$, the situation is reversed for trace $\mathcal{C}$. To understand this, we observed that most missed DDoS attacks are mostly originating with or targeting a random port number. In this paper, we have adopted the rule $(srcAS \wedge dstAS \wedge srcPort \wedge dstPort)$ in flagging anomalous time bins. This rule could not be best suitable for detecting such anomalies and new rules could be tried, but we leave this for future investigation. 

\begin{table}
\caption{Anomalies found by each approach: SENATUS H1, H2 and Apriori (AP)}
\label{anomaliespertype}
\centering
\begin{tabular}{|c c c c c|}
\hline
 & & Trace $\mathcal{A}$& &\\
 \hline
\end{tabular}
\begin{tabular}{|c|c|c|c|c|}
\hline
Anomaly type & Total & $\mathcal{H}$1 &$\mathcal{H}2$ &$AP$\\
\hline
\hline
DDoS & 75& 33& 10&32\\ 
\hline
DoS& 12& 4& 7& 1\\
\hline
Scans &223& 58& 96& 69\\
\hline
Total &310 &95& 113& 102\\
\hline
\end{tabular}
\begin{tabular}{|c c c c c|}
\hline
 & & Trace $\mathcal{B}$& &\\
 \hline
\end{tabular}
\begin{tabular}{|c|c|c|c|c|}
\hline
Anomaly type & Total & $\mathcal{H}$1 &$\mathcal{H}2$ &$AP$\\
\hline
DDoS & 76& 34& 15&27\\
\hline
DoS& 18& 8& 8& 2\\
\hline
Scans &335& 125& 167& 43\\
\hline
Total &429 &167& 190& 72\\
\hline
\end{tabular}
\begin{tabular}{|c c c c c|}
\hline
 & & Trace $\mathcal{C}$& &\\
 \hline
\end{tabular}
\begin{tabular}{|c|c|c|c|c|}
\hline
Anomaly type & Total & $\mathcal{H}$1 &$\mathcal{H}2$ &$AP$\\
\hline
\hline
DDoS & 54& 16& 13&25\\
\hline
DoS& 22& 5& 3& 14\\
\hline
Scans &374& 153& 178& 43\\
\hline
Total &450 &174& 194& 82\\
\hline
\end{tabular}
\begin{tabular}{|c c c c c|}
\hline
 & & Trace $\mathcal{D}$& &\\
 \hline
\end{tabular}
\begin{tabular}{|c|c|c|c|c|}
\cline{1-5}
Anomaly type & Total & $\mathcal{H}$1 &$\mathcal{H}2$ &$AP$\\
\hline
\hline
DDoS & 101& 24& 18&59\\
\hline
DoS& 10& 2& 4& 4\\
\hline
Scans &220& 75& 118& 27\\
\hline
Total &331 &101& 140& 90\\
\hline
\end{tabular}
\end{table}

In addition, Table \ref{anomaliespertype} shows that SENATUS using H1 finds more DDoS attacks than using H2. This is likely due to the fact that only one third (around $29\%$) of the DDoS attacks found in the collected dataset have packets size less than $64$ bytes, as indicated by Figure~\ref{anomdistributions}.

Furthermore, we have observed that SENATUS using H2 detects more network scans than using H1. Most of the additionally detected network scans are small intensity SYN scans using small packets, which are more easily spotted using the second heuristic.

\subsection{Performance Comparison}

We now evaluate the tradeoff between detection and false positive rates for SENATUS(H1 $\bigcup$ H2), SENATUS(H1), SENATUS(H2) and Apriori. Here, the true positive rate is defined as the ratio of the detected number of anomalies using the method with respect to the total detected number of anomalies using either method. In addition, the false positive rate is defined as the ratio of the number of false positives caused by the method with respect to the number of detected anomalies using this method. For ease of notation, we refer the true positive rate of the combined set of anomalies resulting from the union of H1 and H2, i.e. SENATUS(H1 $\bigcup$ H2), as SENATUS' true positive rate. As in calculating the false positive rate for SENATUS(H1 $\bigcup$ H2), if any of SENATUS(H1) and SENATUS(H2) flags an anomaly but it is identified by visual inspection as a false positive, then the combined number of false positive for SENATUS(H1 $\bigcup$ H2) increments by one.
\begin{figure}
\centering
\subfigure{
   \includegraphics[width=0.22\textwidth]{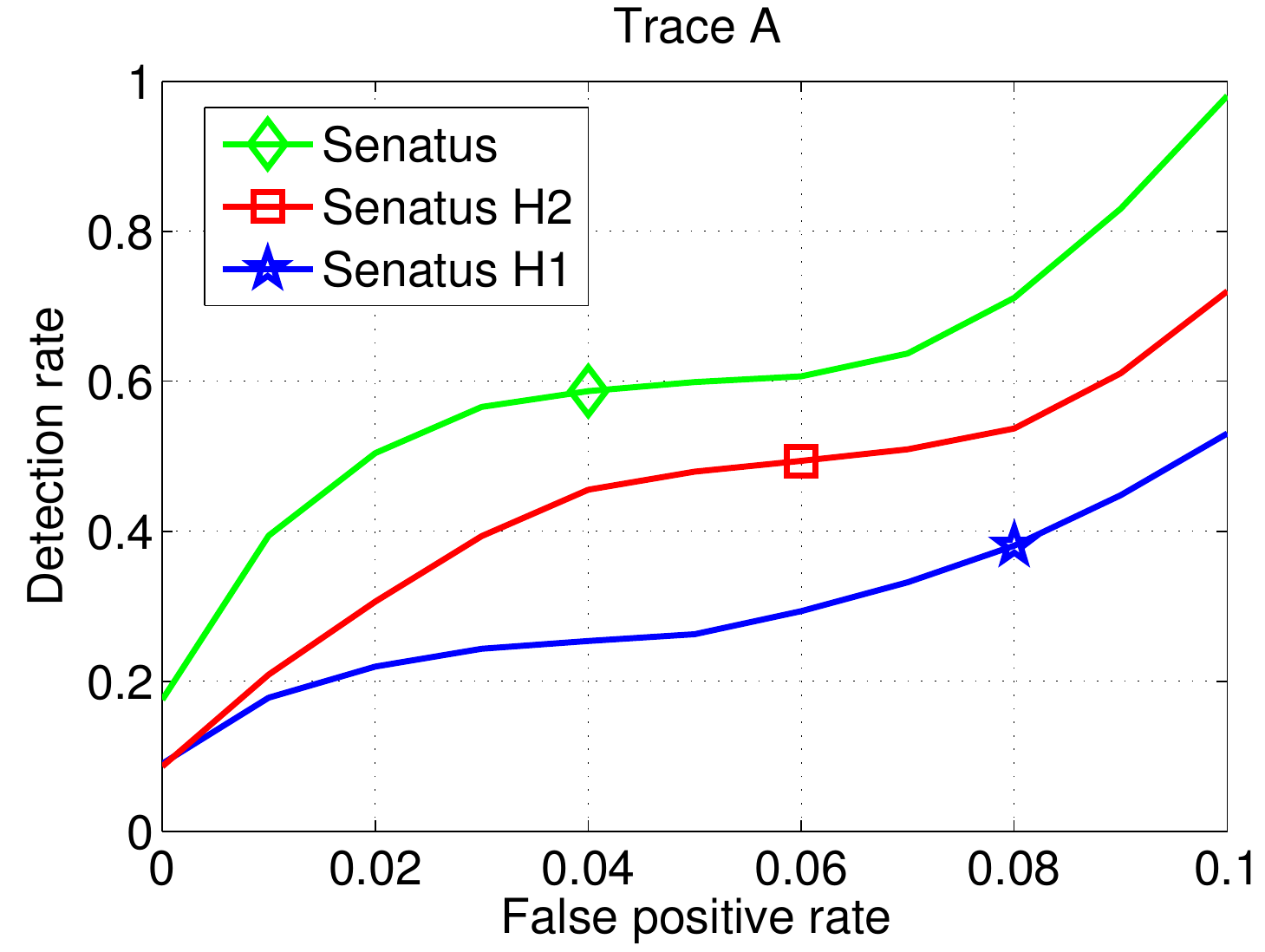}
}
\subfigure{
   \includegraphics[width=0.22\textwidth]{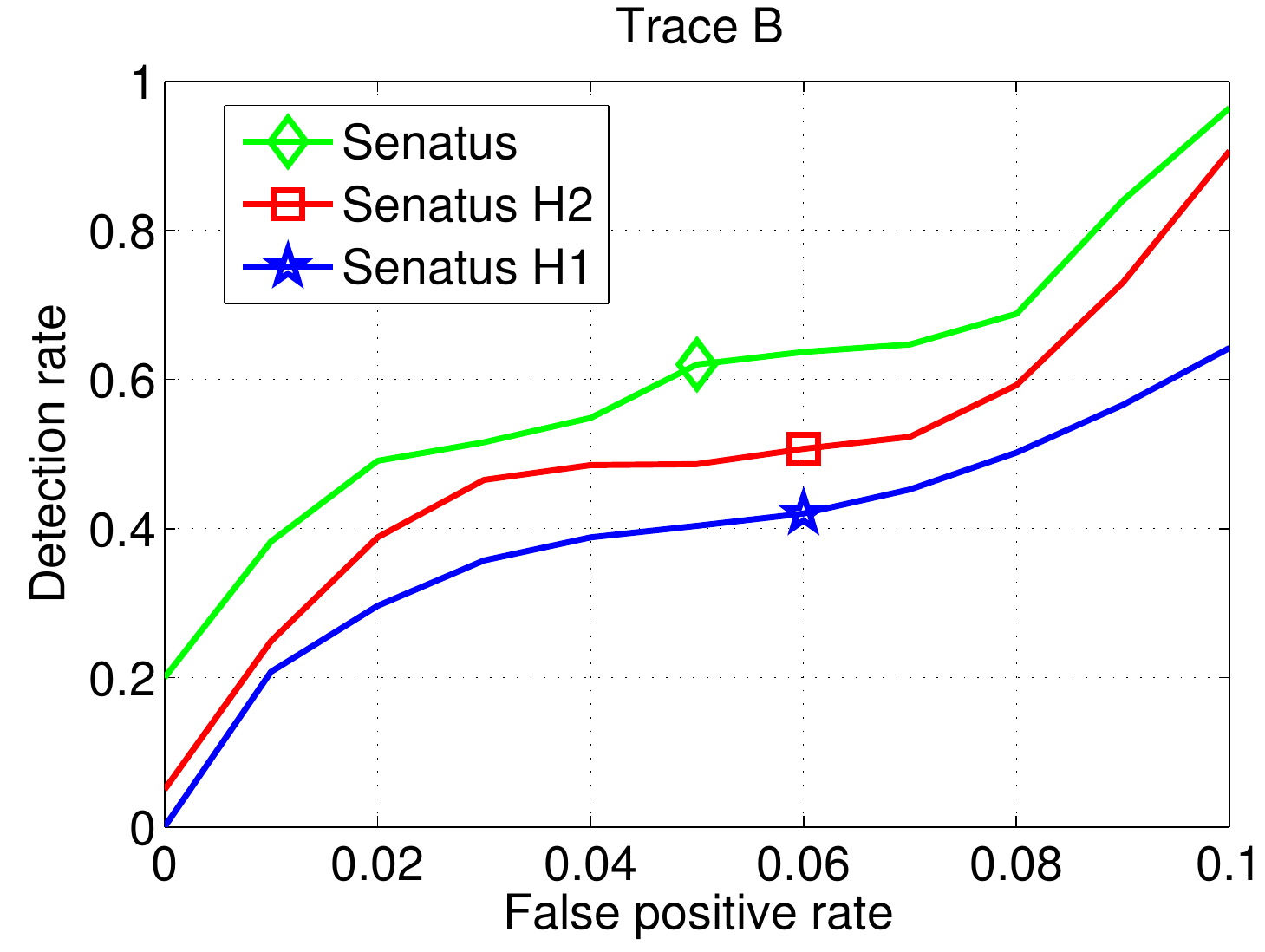}
}
\subfigure{
   \includegraphics[width=0.22\textwidth]{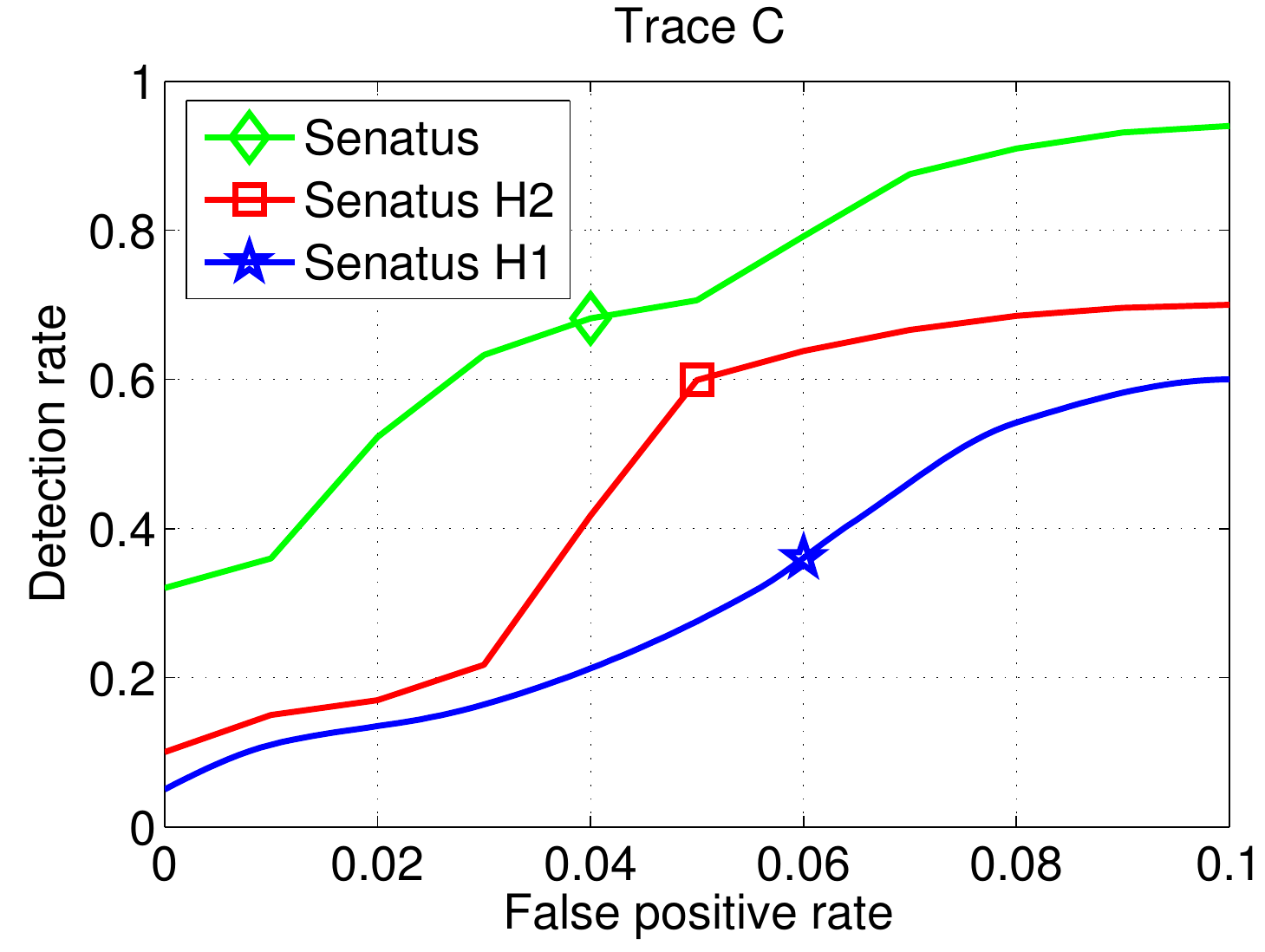}
}
\subfigure{
  \includegraphics[width=0.22\textwidth]{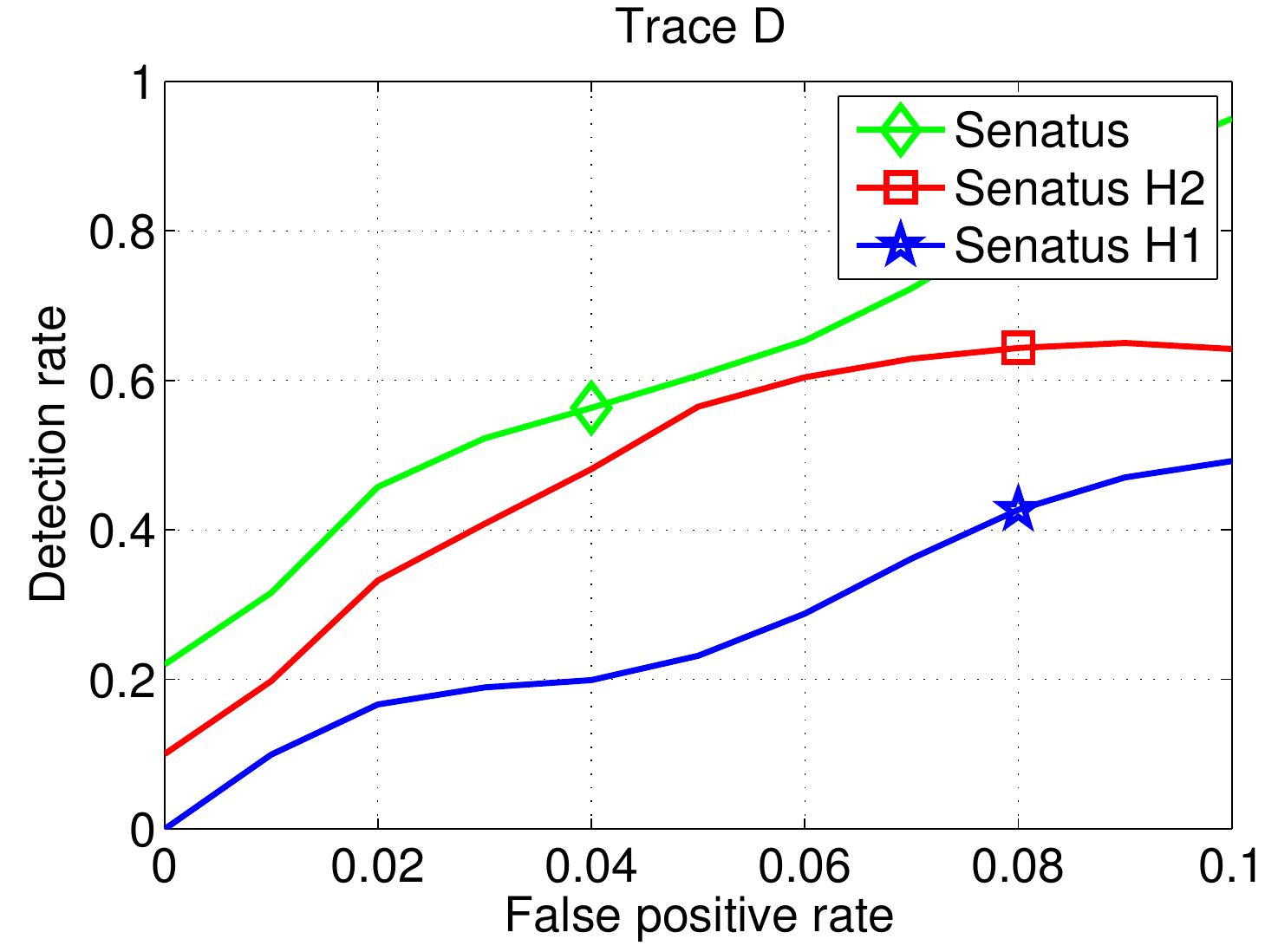}
}
\caption{Receiver Operator Characteristics (ROC) curves}\label{roc}
\vspace{-0.3cm}
\end{figure}

Figure \ref{roc} displays the ROC (Receiver Operator Characteristics) curves that illustrate the detection rate of SENATUS as a function of the false positive rate. The figure firstly shows that while the detection rate of SENATUS using both H1 or H2 is low, the detection rate of SENATUS using the union of the two heuristics is much higher. For example, the detection rate for SENATUS(H2), SENATUS(H1), and SENATUS (H1 $\bigcup$ H2) are around $73\%$, around $64\%$, and $84\%$ in trace $\mathcal{B}$. The figure additionally shows that the false positive rate is low for all collected traces: it does not exceed $10\%$ for the four collected traces.

Interestingly, the figure highlights the findings previously discussed: the detection rate of SENATUS using either H1 or H2 is low compared to using the union H1 $\bigcup$ H2. Using the union\nop{ of the two heuristics} to construct the traffic population, further analyzed by SENATUS, helps to improve the overall detection rate due to the {\em complimentary nature} of the two heuristics. 
\begin{figure}
  \centering
    \includegraphics[width=0.4\textwidth]{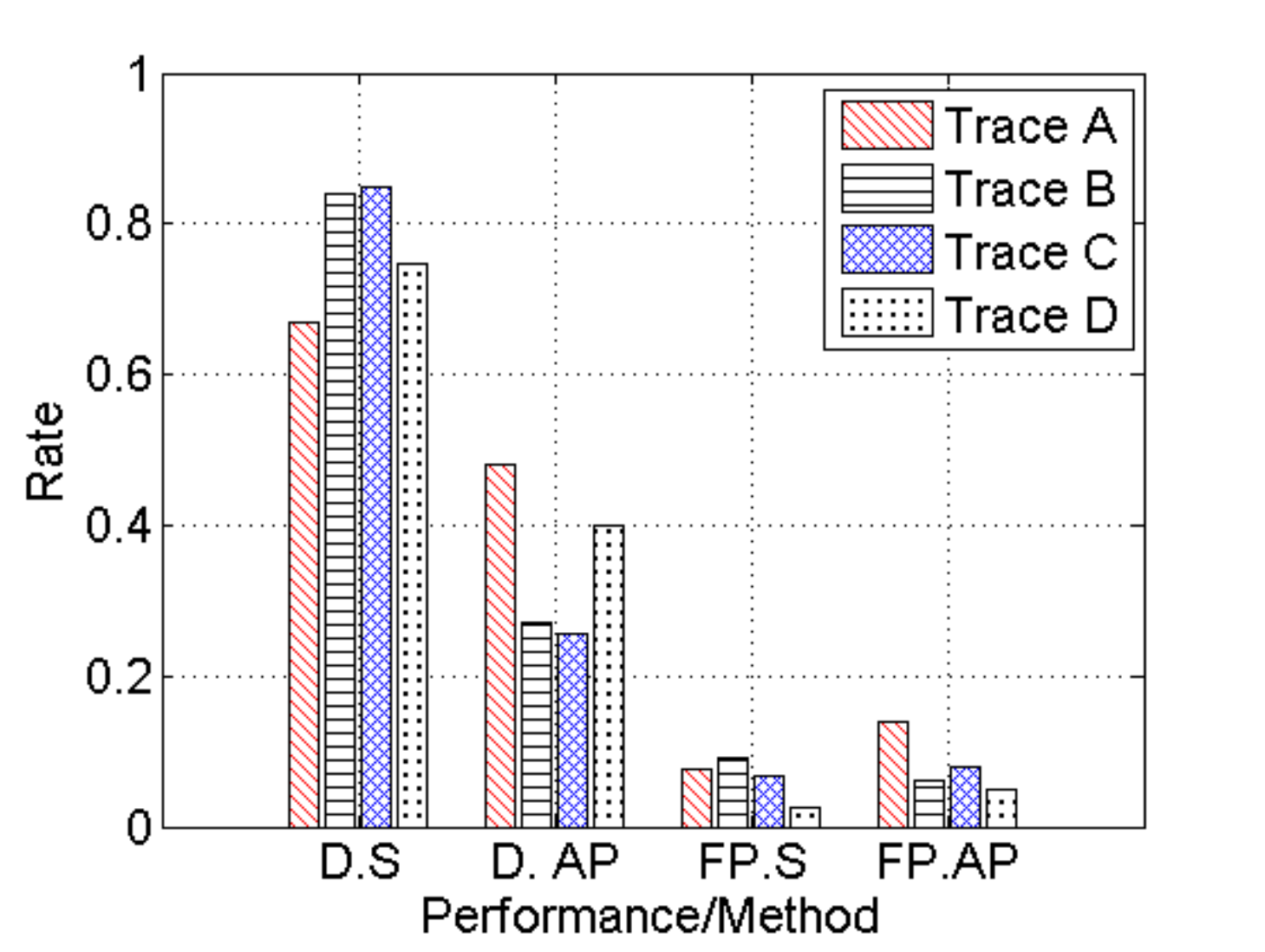}
  \caption{D.{\em X}: true positive rate of {\em X}, and FP.{\em X}: false positive rate of {\em X}, where {\em X}= S for SENATUS and {\em X}= AP for Apriori.}\label{comp}
  \vspace{-0.5cm}
\end{figure}

To illustrate more directly the detection-false positive tradeoff for SENATUS (H1 $\bigcup$ H2) and Apriori, Figure \ref{comp} is presented. Each bar graph shows either the detection or the false positive rate for SENATUS or Apriori  and for each of the collected four traces.\nop{ D, FP, S and AP in the figure denote true positive rate, false positive rate, SENATUS, Apriori respectively.} The figure shows that while SENATUS experiences the best detection-false positive tradeoff, Apriori generally exhibits the lowest detection and the highest false positive rate among the three approaches for the four collected traces. For example, for trace $D$, while the SENATUS' detection rate is about 75\% with a false positive rate about 2\%, the Apriori's detection rate is around 40\% with a false positive rate about 5\%.

\section{Related work} \label{sec-6}
The problem of network anomaly detection has attracted a lot of research effort. This section discusses the works related to\nop{ our proposed SENATUS approach} SENATUS. These works are divided into two categories: traffic anomaly detection and root cause analysis.

\subsection{Traffic Anomaly Detection} 

Significant attention has been devoted to developing various traffic anomaly detection techniques. Earlier techniques have mostly relied on volume metrics such as packets, bytes and flows for analysis using time series prediction \cite{soule05}, signal processing \cite{barford02} or machine learning \cite{sequeira02}, in order to detect abrupt variations in traffic volume signals. In \cite{newwork1}, Kullback-Leibler (KL) divergence-based method is proposed for detecting anomalous traffic mimicking legitimate traffic, which is similar to the Apriori approach. In \cite{lakhina05}, Lakhina et al  proposed an approach that relies on traffic histograms (traffic distributions) for anomaly detection, motivated by the observation that anomalies distort the distribution of the traffic over feature values e.g. IP addresses and ports. Unfortunately, traffic histograms suffer from the curse of dimensionality issue. To deal with this challenge, Lakhina et al proposed to use entropy to summarize a traffic histogram into one value \cite{lakhina05}. 

The entropy method has been widely adopted for traffic anomaly detection. In \cite{Mon2015}, anomaly detection in context of detecting DDoS is experimented with four important information entropy measures: Hartley entropy, Shannon entropy, Renyi's entropy and Generalized entropy. It has been observed that the use of an appropriate information metric helps to magnify the spacing between legitimate and attack traffic for both low-rate and high-rate DDoS attack detection in real network traffic. Recently in \cite{yan2016}, the entropy theory and support vector machine have been used to detect network anomaly traffic. 

However, the entropy method was shown to only coarsely model the properties of traffic histograms thus ineffective to detect a wide range of traffic anomalies \cite{kind09}. On the other hand, other histogram-based anomaly detection approaches have faced the challenge of traffic histogram dimensionality reduction \cite{ramah09, kind09, brauckoff09imc}. To this end, the authors of \cite{kind09} proposed to keep the well-known source and destination ports, remove the components that remain constant, which are associated to unused feature values, and additionally apply the Principal Component Analysis (PCA) technique. In addition, the authors of \cite{brauckoff09} proposed an aggregation strategy using hash functions to reduce traffic histograms dimension. Their approach was shown to be promising, providing a lossless compression technique for traffic histogram analysis. However, it suffers from a serious weakness: a map between the hash function and the original histogram is required, which adds an additional non-negligible processing overhead. In \cite{mardani13}, network anomalies are detected via sparsity and low rank property. There, the goal was to construct a map of anomalies in real time, that summarizes the network ``health state'' along both the flow and time dimensions. Recently, another anomaly detection scheme was proposed in \cite{guha}, which focuses on the anomaly detection problem for dynamic data streams through the lens of random cut forests.

SENATUS proposes a traffic anomaly detection technique that also relies on traffic histograms. Differently, SENATUS deals with the curse of dimensionality using a simple low-complexity lossy compression approach\nop{ that was recently proposed}, which only extracts the top-$K$ components of the histograms \cite{atef12}. 

\subsection{Root-Cause Analysis} 

In the literature, very few works have tried to address the problem of root-cause analysis of the traffic anomalies related alarms, which typically consists on identifying the flows involved in the anomalous behavior and pinpointing the anomalies causing the identified behavior.

Fernandes et al \cite{fernandes09} tried to address this problem using a set of predefined signatures, such as traffic descriptors describing the behavior of network attacks including DoS, DDoS and scans. These predefined signatures involve a large number of empirical threshold values. To circumvent the issues with signature-based anomaly classification, in \cite{silvera10}, manually identified anomalies were projected on a 22-coordinate feature space to classify new anomalous behavior through hierarchical clustering. However, if a new behavior, which is different from the built-in anomalies, is encountered, the used methodology is unable to classify or characterize the considered unknown anomaly. To deal with this challenge, the authors of \cite{casas2012} have proposed an unsupervised algorithm which combines the notions of Sub-Space Clustering and Evidence Accumulation clustering on 36 two-dimensional feature subspace. This approach may suffer from scalability issues when implemented on high speed links in a backbone network. Moreover, the authors of \cite{paredes12} proposed a data mining approach to classify detected anomalous flows. This approach focuses on finding flows, in the data set, sharing feature values that appear frequently together and the frequency is greater than a predefined threshold called the minimum support. While it was proven to be efficient to classify a wide class of anomalies, the minimum support threshold is generally difficult to choose \cite{paredes12}.
In \cite{silva2016}, a framework is proposed, called as ATLANTIC, which combines the use of information theory to calculate deviations in the entropy of flow tables and a range of machine learning algorithms to classify traffic flows. 
In addition, authors of \cite{wang2015} proposed a technique for root cause analysis in component-based systems and their approach focuses at application-level anomaly correlation.

SENATUS, on the other hand, relies on a \textbf{linear time} algorithm where the threshold values are automatically identified based on a machine learning classification technique, the decision tree method (Random Tree) \cite{witten11} only using four-tuple features: srcIP, dstIP, srcPort and dstPort.

\section{Conclusion} \label{sec-7}
In this paper, we propose SENATUS, a novel approach for traffic anomaly detection and root-cause analysis. In particular, it conducts root cause analysis jointly with anomaly detection. In addition to the novel joint treatment, the specific novelty and contribution of SENATUS are as follows. (1) First, instead of performing analysis directly on the original traffic histogram which suffers from the curse of dimensionality, we propose to use approximate traffic histograms with much reduced dimensionality as inputs to the analysis. Conceptually, this forms the core of the SENATUS election stage. (2) Second, at the SENATUS voting stage, we propose to use PCP to detect time bins with abrupt changes in the time series of each selected traffic feature value. Detected abrupt variations serve as votes which flag if a time bin is anomalous using a defined decision rule. (3) At the final decision stage, we further conduct root-cause analysis on each anomalous time bin. We propose to adopt a machine learning-based technique for this purpose. (4) For the GEANT measurement dataset of four 18 day-long traces, we evaluate SENATUS and compare with a well-known traffic histogram-based anomaly detector, Apriori. We found that SENATUS uncovers a high number of anomalies that Apriori does not, in addition to SENATUS' better performance in diagnosing network scans and DoS/DDoS for the GEANT dataset. This makes SENATUS an appealing approach for joint traffic anomaly detection and root-cause analysis, well complimenting the Apriori approach.

\nop{

In this paper, we proposed SENATUS, a novel approach for traffic anomaly detection and root-cause analysis. In particular, it conducts root cause analysis jointly with anomaly detection. In addition to the novel joint treatment, the specific novelty and contribution of SENATUS are as follows. (1) First, instead of performing analysis directly on the original traffic histogram which suffers from the curse of dimensionality, we propose to use aggregate and approximate pre-filtered traffic histograms with much reduced dimensionality as inputs to the analysis. Conceptually, this forms the core of the SENATUS election stage, where each chosen component is treated as a senator. (2) Second, at the SENATUS voting stage, we propose to use PCP to detect time bins with abrupt changes in the time series of each selected traffic feature value (i.e. senator). Detected abrupt variations serve as votes which flag anomalous time bin using a defined decision rule. (3) A final decision stage where we further conduct a root-cause analysis on each anomalous time bin. We propose to adopt a linear time signature in addition to a machine learning-based technique for this purpose. (4) For the 18 days-long four GEANT2 measurement dataset, we evaluate SENATUS and compare with a similar detector relying on traffic histograms analysis using lossless compression, called Apriori. We found that in addition to its good performance in diagnosing network scans and DoS/DDoS attacks, SENATUS uncovers a high number of anomalies that Apriori does not.
}

\section*{Acknowledgment}  
\noindent This research was partly funded by the EU FP7 Marie Curie Actions Cleansky Project, Contract No. 607584. 
\bibliographystyle{ieeetr}
\bibliography{PhDmidTerm} 

\end{document}